\documentclass[dvipdfm,conference]{IEEEtran}
\IEEEoverridecommandlockouts

\usepackage{cite}
\usepackage{amsmath,amssymb,amsfonts}
\usepackage{algorithmic}

\usepackage{textcomp}
\def\BibTeX{{\rm B\kern-.05em{\sc i\kern-.025em b}\kern-.08em
    T\kern-.1667em\lower.7ex\hbox{E}\kern-.125emX}}
\usepackage{tabularx}
\usepackage{graphicx}
\usepackage{wrapfig}
\usepackage{epstopdf}

\newcommand{\vc}{\mathrm{vec}}

\newcommand{\diag}{\mathrm{diag}}

\newcommand{\snr}{\small \mbox{SNR}}

\newcommand{\bPhi}{\boldsymbol{\Phi}}

\newcommand{\bphi}{\boldsymbol{\phi}}
\newcommand{\bpsi}{\boldsymbol{\psi}}

\newcommand{\bA}{\boldsymbol{A}}

\newcommand{\bF}{\boldsymbol{F}}
\newcommand{\bG}{\boldsymbol{G}}
\newcommand{\bH}{\boldsymbol{H}}
\newcommand{\bI}{\boldsymbol{I}}

\newcommand{\bK}{\boldsymbol{K}}

\newcommand{\bM}{\boldsymbol{M}}

\newcommand{\ba}{\boldsymbol{a}}

\newcommand{\bff}{\boldsymbol{f}}
\newcommand{\bg}{\boldsymbol{g}}
\newcommand{\bh}{\boldsymbol{h}}

\newcommand{\bu}{\boldsymbol{u}}
\newcommand{\bv}{\boldsymbol{v}}
\newcommand{\bw}{\boldsymbol{w}}
\newcommand{\bx}{\boldsymbol{x}}
\newcommand{\by}{\boldsymbol{y}}
\newcommand{\bz}{\boldsymbol{z}}

\newcommand{\bzero}{\boldsymbol{0}}

\newcommand{\bVb}{{\tilde \bVb}}

\newcommand{\cCN}{\mathcal{CN}}

\newcommand{\tr}{\mathrm{tr}}

\newcommand{\T}{\tiny \mbox{T}}  
\newcommand{\R}{\tiny \mbox{R}}

\newcommand{\nt}{n_{\T}}
\newcommand{\nr}{n_{\R}}
\newcommand{\IS}{\tiny \mbox{IS}}  
\newcommand{\nis}{n_{\IS}}
\newcommand{\K}{\tiny \mbox{K}}
\newcommand{\M}{\tiny \mbox{M}}
\newcommand{\bgtil}{{\tilde \bg}}
\newcommand{\bhtil}{{\tilde \bh}}
\newcommand{\HH}{\tiny \mbox{H}}  
\newcommand{\GG}{\tiny \mbox{G}}

\begin{document}

\title{Optimization of Reconfigurable Intelligent Surfaces Through Trace Maximization
\thanks{Sayeed's work was partly supported by the US NSF through grants \#1703389 and \#1629713.}}

\author{\IEEEauthorblockN{Akbar M. Sayeed}
\IEEEauthorblockA{\textit{Department of Electrical and Computer Engineering}\\
\textit{University of Wisconsin}\\
\textit{Madison, WI 53711}\\
 akbar.sayeed@wisc.edu}}

\maketitle

\begin{abstract}
Reconfigurable Intelligent Surfaces (RIS) have received significant attention recently as an innovation for enhanced connectivity, capacity, and energy efficiency in future wireless networks.  Recent works indicate that such RIS-augmented communications can significantly enhance performance by intelligently shaping the characteristics of the multipath propagation environment to focus the energy in a desired direction and to circumvent impediments such as blockage, especially for communication at millimeter-wave (mmW), Terahertz (THz) and higher frequencies. In this paper, we investigate optimized (amplitude and phase) RIS design in a point-to-point multipath MIMO link and study the impact on link capacity under the assumption of perfect channel state information at the transmitter (TX), receiver (RX) and RIS. Specifically, we propose RIS design based on the maximization of the trace of the composite TX-RIS-RX link matrix which is a measure of the average power at the RX. We propose two RIS designs: a diagonal RIS matrix, and a general RIS matrix representing a more advanced architecture. The optimum design, in both cases, corresponds to calculating the dominant eigenvector of certain Hermitian matrices induced by the component channel matrices. We illustrate the capacity performance of the optimized RIS designs and compare them to a baseline design (random amplitudes and phases) and a recently proposed low-complexity phase-only design. We present results for sparse and rich multipath, and also consider the impact of line-of-sight paths. Our results show that while all designs offer comparable capacity at high signal-to-noise ratios (SNRs), the proposed optimum designs offer substantial gains at lower SNRs.     
\end{abstract}

\section{Introduction}
\label{sec:intro}
Reconfigurable Intelligent Surfaces (RIS) are the focus on significant current research for enhancing the connectivity, capacity, and energy efficiency of future wireless networks; see, e.g., \cite{ris_alouini1:19, ris_alouini2:20, ris_fink:19, ris_gli:20, ris_zhang1:20} for different overview perspectives in the area, including a discussion of research challenges.  
At a basic level, an RIS can be viewed as an engineered and reconfigurable surface that can shape the amplitude, phase, polarization and/or frequency of the incident electromagnetic (EM) wavefront and then re-emitting the modified wavefront to enhance the end-to-end performance between a transmitter (TX) and receiver (RX) \cite{ris_alouini1:19}. Similarly, a collection of strategically placed RISs can augment the performance of a multi-user network \cite{ris_zhang1:20}.  The interest in RIS-aided communication is partly due to the advancements and innovations in the design of EM metasurfaces that manipulate the properties of the EM wavefronts \cite{ris_fink:19}, analogous to spatial light modulators. Another motivation for RIS-aided communication is to circumvent impediments to propagation at mmW, THz and higher frequencies, such as blockage and attenuated reflections, due to the highly directional, quasi-optical nature of communication at such frequencies. Ongoing work by a number of researchers is exploring various aspects of RIS design, including EM and hardware considerations \cite{ris_alouini2:20, ris_kkwong:20}, energy and algorithmic efficiency \cite{ris_debbah:19, ris_alouini1:19}, multiuser and MIMO communication \cite{ris_zhang2:20, ris_kkwong:20}, and new communication-theoretic formulations \cite{ris_fink:19}.  

\begin{figure}[htb]
\vspace{-3mm}
\centerline{ \includegraphics[width=0.40\textwidth]{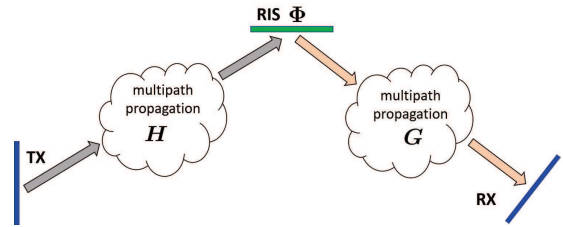}}
\caption{\footnotesize{\sl A schematic illustrating the system model considered in this paper. A TX communicates with a RX through an RIS. It is assumed that there are no direct propagation paths connecting the TX with the RX.}}
\label{fig:ris}
\end{figure}

In this paper, we consider optimized design of an RIS, with configurable amplitudes and phases, in the simplest setting of a point-to-point multiple input multiple output (MIMO) link mediated by an RIS, as illustrated in Fig.~\ref{fig:ris}. The TX to RIS channel is denoted by the matrix $\bH$ and the RIS to RX channel is denoted by the matrix $\bG$. The action of the RIS is modeled by the matrix $\bPhi$ resulting in the composite TX-RIS-RX channel matrix $\bF = \bG \bPhi \bH$.  To focus on the impact of $\bPhi$, we assume that there are no direct propagation paths coupling the TX and the RX - all paths are through the RIS. We consider physically meaningful multipath models for both $\bH$ and $\bG$ and consider the impact of both line-of-sight (LoS) and non-LoS (NLoS) propagation. The basis system model, including physical models for $\bH$ and $\bG$,  is developed in Sec.~\ref{sec:model}.

We assume noiseless operation of the RIS
\begin{equation}
\by_o = \bPhi \by_i 
\label{ris_io}
\end{equation} 
where $\by_i$ denotes the signal vector incident on the RIS elements and $\by_o$ denotes the signal vector emitted by the RIS.  Typically, $\bPhi$ is assumed to be diagonal matrix, representing an element-wise manipulation of amplitude and phases of the incident EM wave. By appropriately choosing $\bPhi$, we can model different scenarios. For example,  a diagonal $\bPhi$ with random complex entries can model the random reflections from a passive physical object and serves as a baseline (RAND) model for non-engineered surfaces. We also consider a recently proposed low-complexity phase-only (LC-PH) design for RIS-aided MIMO links  \cite{ris_lc:20}.

We propose a framework for optimized $\bPhi$ design based on maximization of trace (the sum of the diagonal matrix entries): 
\begin{equation}
\sigma^2_{\bF} = \tr(\bF^\dagger \bF)  
\label{tr_opt}
\end{equation}
which, as we discuss later, is a measure of the average power at the RX.  We consider two optimized designs: i) diagonal $\bPhi$ (OPT-DIAG), and ii) general $\bPhi$ (OPT-GEN). We note that the general $\bPhi$ design represents a more advanced RIS in which the incoming signals on the RIS elements are jointly and linearly processed by the RIS, as in (\ref{ris_io}), before being re-emitted.  
We impose a trace constraint on $\bPhi$ for all choices of $\bPhi$ for a physically meaningful and fair comparison between the performance of various designs (RAND, LC-PH, OPT, OPT-GEN).  The optimum design of $\bPhi$ in both cases is characterized by the dominant eigenvector of certain Hermitian matrices induced by the component channel matrices $\bH$ and $\bG$. 
The optimized design of $\bPhi$, along with a description of the RAND and LC-PH designs, is discussed in Sec.~\ref{sec:opt}.  

In Sec.~\ref{sec:res} we compare the link capacities of various $\bPhi$ designs in rich and sparse multipath environments and also consider the impact of LoS propagation paths and the size of the RIS. Our results indicate that while all designs offer comparable performance at high SNRs, the optimum designs offer a significant gain in capacity at lower SNRs. In particular, the general optimum design has the highest capacity gain at low SNRs but it comes at the cost of loss in capacity compared to other designs at high SNRs. On the other hand, the optimum diagonal $\bPhi$ design offers the best performance over the entire SNR range - higher capacity at lower SNRs and comparable capacity at high SNRs relative to the baseline (RAND) design and the low-complexity phase-only (LC-PH) design. 
Furthermore, our results indicate that we can simply replace the entries of the optimized $\bPhi$ with their phases (while keeping amplitudes constant) without incurring any loss in performance. This is an attractive property from the viewpoint of practical realization of the optimized RIS designs.

The development of the RIS design framework in this paper is guided by fundamental principles of MIMO channel modeling and communication over multipath channels \cite{sayeed:02a}. These principles are also used to provide an interpretation of the results. One defining feature of the trace-optimized designs proposed in this paper is that they concentrate the channel power in a small number of eigenvalues and this concentration is most pronounced for the general $\bPhi$ design.  The framework laid out in this paper opens new directions for research and innovation in the design, optimization and implementation of RISs which are briefly discussed in Sec.~\ref{sec:conc}.

{\bf Notation:} Lowercase boldfaced letters represent (column) vectors and uppercase boldfaced letters represent matrices.  The complex conjugate (Hermitian) transpose of $\bA$ is denoted by $\bA^\dagger$,  $\bA^{\T}$ denotes a simple transpose, and $\bA^*$ denotes complex conjugation; that is$\bA^\dagger = \bA^{\T*}$. 

\section{System Model}
\label{sec:model}
For simplicity of exposition, we develop the basic ideas in the context of the TX,  the RX, and the RIS consisting of critically (half wavelength) spaced uniform linear arrays (ULAs) of elements. The basis architecture is illustrated in Fig.~\ref{fig:ris}.  A TX with an $\nt$-element ULA is communicating with a RX with an $\nr$-element ULA through an $\nis$-element RIS.   The TX-RIS channel is denoted by the $\nis \times \nt$ matrix $\bH$ and the RIS-RX channel is denoted by the $\nr \times \nis$ matrix $\bG$. The action of the RIS is represented by the $\nis \times \nis$ matrix $\bPhi$. To focus on the impact of the RIS on the overall TX-RIS-RX link, we assume that there are no direct propagation paths connecting the TX and the RX.  By appropriately choosing $\bPhi$ we can model the effect of random uncontrolled connections between the incoming and outgoing signals at the RIS, as may be the case of a passive physical object (e.g., a concrete wall) occupying the same physical space as the RIS. On the other hand, by explicitly designing $\bPhi$,  based on the available channel state information (CSI) about $\bH$ and $\bG$, we can model the impact of an engineered RIS. The critically spaced elements in the RIS enable this comparison without any loss of signal information \cite{sayeed:02a,sayeed:handbook08}.  Furthermore, we focus on the spatial propagation aspects and thus consider non-selective channel impact over the bandwidth and duration of signal transmission.

Specifically, the model for the TX-RIS-RX system in Fig.~\ref{fig:ris} can be expressed as
\begin{equation}
\bz = \bG \bPhi \bH \bx + \bw = \bF \bx + \bw
\label{sys_mod}
\end{equation}
where $\bx$ is the $\nt$-dimensional transmitted signal vector, $\bz$ is the $\nr$-dimensional received  signal, $\bw \sim \cCN(\bzero, \bI_{\nr})$ is the complex additive Gaussian noise (we ignore any noise added at the RIS in this paper), and $\bF = \bG \bPhi \bH$ is the $\nr \times \nt$ composite channel matrix representing the overall TX-RIS-RX link.  To make the signals at the RIS more explicit, the incoming signal at the RIS is given  $\by_i = \bH \bx$,  the outgoing signal at the RIS is $\by_o = \bPhi \by_i = \bPhi \bH \bx$ and the signal at the RX is $\bz = \bG \by_o  + \bw = \bF \bx + \bw$. 

\subsection{Modeling of RIS Matrix $\bPhi$}
\label{sec:Phi}
In most cases of interest $\bPhi$ can be modeled as a diagonal matrix. For modeling an un-engineered RIS, $\bPhi$ could be  modeled as $\bPhi = \bI_{\nis}$ or consisting of random phases and amplitudes. For an engineered RIS, the diagonal entries of $\bPhi$ are chosen based on CSI: $\bPhi = \bPhi(\bG,\bH)$.  We assume perfect knowledge of $\bH$ and $\bG$ in the design of $\bPhi$.  For fair comparison, we impose a power constraint on $\bPhi$
\begin{equation}
\tr(\bPhi^\dagger \bPhi) = \nis
\label{ris_const}
\end{equation}
which holds  if all the diagonal entries of $\bPhi$  have unit magnitude (phase-only), for example. For the optimized designs, we assume the general case of amplitude and phase control of the entries of $\bPhi$ while satisfying the power constraint (\ref{ris_const}).

\subsection{Modeling of Channel Matrices $\bH$ and $\bG$}
\label{sec:chan}
For a realistic evaluation of the proposed  designs for $\bPhi$, we consider physical models (rather than purely statistical models, e.g., i.i.d. fading) for $\bH$ and $\bG$ representing a multipath propagation environment  \cite{sayeed:02a, sayeed:handbook08}:
\begin{align}
\bH & = \sum_{\ell = 1}^{N_{p}^{\HH}} \alpha^{\HH}_\ell \ba_{\nis}(\theta^{\HH}_{\R,\ell}) \ba_{\nt}^\dagger (\theta^{\HH}_{\T,\ell}) \label{Hmod} \\
\bG & = \sum_{\ell = 1}^{N_{p}^{\GG}} \alpha^{\GG}_\ell \ba_{\nr}(\theta^{\GG}_{\R,\ell}) \ba_{\nis}^\dagger(\theta^{\GG}_{\T,\ell})  \label{Gmod} 
\end{align}
where $N_p$ denotes the number of paths,  $\alpha_\ell $ denotes the complex path gain, and $\theta_{\R,\ell}$ and $\theta_{\T,\ell}$ denote the spatial frequencies  corresponding to the angle of arrival (AoA) and angle of departure (AoD), respectively, for the $\ell$-th path. For half-wavelength spaced elements the spatial frequency is related to the angle $\varphi$ as $\theta = 0.5 \sin(\varphi)$. The vectors $\ba_{\nr}(\theta)$, $\ba_{\nt}(\theta)$, and $\ba_{\nis}(\theta)$ denote the array response/steering vectors for the ULAs at the RX, TX and RIS, respectively:
\begin{equation}
\ba_n(\theta) = \left [1, e^{j2\pi \theta}, e^{j4\pi \theta}, \cdots, e^{j2\pi(n-1)\theta} \right ]^{\T}  \ .
\label{ba}
\end{equation}   
The spatial frequencies are randomly distributed over the maximum angular spreads $\theta_\ell \in [-0.5,0.5]$ and the path gains are statistically independent and distributed as $\alpha_\ell \sim \cCN(0,\sigma^2_\ell)$ for NLoS paths and only consist of random phase (with appropriately scaled amplitude) for LoS paths.  

\section{Optimized Design of RIS matrix $\bPhi$}
\label{sec:opt}
In this section, we present the main results on the proposed design for $\bPhi$. Our approach is based on maximizing the power of the composite channel matrix $\bF = \bH \bPhi \bG$. Formally, the design problem can be stated as 
\begin{equation}
\bPhi_{opt} = \arg \max_{\Phi} \tr(\bF^\dagger \bF)  \  \mbox{subject  to\ } \tr(\bPhi^\dagger \bPhi) = \nis
\label{opt_Phi}
\end{equation}
The motivation for maximizing the trace of $\bF^\dagger \bF$ comes from the observation that it is a measure of the total average power at the RX. Conditioned on a particular $\bF$, the average power in the received signal vector $\bz$ in (\ref{sys_mod}) is
\begin{align}
E[|\| \bz \|^2] & =  E [ \bx^\dagger \bF^\dagger \bF \bx]  + E[\|\bw\|^2]   \nonumber \\ 
& = \tr(\bF^\dagger \bF E [ \bx \bx^\dagger]) + \nr  \nonumber \\
&  =  \tr( \bF^\dagger \bF P_{ave} \bI_{\nt})  + \nr   \nonumber \\
& = P_{ave} \tr(\bF^\dagger \bF) + \nr
\label{rx_pow}
\end{align}
where we have assumed that the transmitted signal vector $\bx$ consists of independent symbols, with average power $P_{ave}$. Thus, maximizing the trace of $\bF^\dagger \bF$ is equivalent to maximizing the RX signal power, which in turn can boost the link capacity and the signal-to-noise ratio (SNR) at the RX.   
\subsection{Design of Diagonal $\bPhi$ (OPT-DIAG)}
\label{sec:diag}
First consider the case of diagonal $\bPhi$ and let $\bphi = [\phi_1, \phi_2, \cdots, \phi_{\nis}]^{\T}$ denote the vector of diagonal entries of $\bPhi$. The composite matrix $\bF$ can be expressed as 
\begin{equation}
\bF =  \bG \bPhi \bH = \sum_{i=1}^{\nis}  \phi_i \bg_i \bh_i^\dagger 
\label{F_phi}
\end{equation} 
where $\bg_i$ denotes the $i$-th column of $\bG$ and $\bh_i^\dagger$ denotes the $i$-th row of $\bH$. The functional in (\ref{opt_Phi})  can be expressed as 
\begin{align}
\tr(\bF^\dagger \bF) & = \sum_{i=1}^{\nis} \sum_{j = 1}^{\nis} \phi_i \phi_j^* \tr(\bh_j \bg_j^\dagger \bg_i \bh_i^\dagger)  \nonumber \\
 & =  \sum_{i=1}^{\nis} \sum_{j = 1}^{\nis} \phi_i \phi_j^* \tr(\bg_j^\dagger \bg_i \bh_i^\dagger \bh_j) \nonumber \\
& =  \sum_{i=1}^{\nis} \sum_{j = 1}^{\nis} \phi_i \phi_j^* \bg_j^\dagger \bg_i \bh_i^\dagger \bh_j \nonumber \\
& =  \sum_{i=1}^{\nis} \sum_{j = 1}^{\nis} \phi_i \phi_j^*  \bK_{ji}  = \bphi^\dagger \bK \bphi  \label{opt_fun} \\
\mbox{where \ } &  \bK_{ji}  = \bg_j^\dagger \bg_i \bh_i^\dagger \bh_j \label{K}
 \end{align}
 Note that $\bK^\dagger = \bK$. The optimum  $\bphi$ directly follows (\ref{opt_fun}).

 \noindent
{\bf Solution for Diagonal} $\bPhi$: The optimum  $\bphi$ solving (\ref{opt_Phi}) for a diagonal $\bPhi$ is a scaled version of the dominant eigenvector, $\bu_{\K}$, corresponding to the largest eigenvalue of $\bK$ in (\ref{K}): 
 \begin{equation}
 \bPhi_{opt} = \diag(\bphi_{opt}) \ ; \ \bphi_{opt} = \sqrt{\nis} \bu_{\K}  \ ; \ \bu_{\K}^\dagger \bu_{\K}  = 1  \ . \label{phi_opt} 
 \end{equation}
 
 
 \subsection{Design of General $\bPhi$ (OPT-GEN)}
 \label{sec:gen}
 Now let us consider the general case of $\bPhi$ that is not constrained to be diagonal.  First, note that 
 \begin{align}
 \bff &= \vc(\bF) = \vc(\bG \bPhi \bH) = [\bH^{\T} \otimes \bG ] \vc(\bPhi) \nonumber \\
   & =   [\bH^{\T} \otimes \bG ] \bpsi \ ; \ \bpsi = \vc(\bPhi)
 \label{Fvec}
 \end{align}
 where $\vc(\cdot)$ denotes the (columnwise)  vectorization of a matrix and $\otimes$ denotes the kronecker product \cite{brewer}.  It follows
 \begin{align}
 \tr(\bF^\dagger \bF) & = \bff^\dagger \bff = \bpsi^\dagger  [\bH^{\T} \otimes \bG ]^\dagger  [\bH^{\T} \otimes \bG ] \bpsi \nonumber \\
  & = \bpsi^\dagger  [\bH^{*} \otimes \bG^\dagger ]  [\bH^{\T} \otimes \bG ] \bpsi \nonumber \\
  & =  \bpsi^\dagger  [\bH^{*} \bH^{\T}  \otimes \bG^\dagger  \bG]   \bpsi \nonumber \\
  & = \bpsi^\dagger \bM \bpsi \ ; \ \bM = [\bH^{*} \bH^{\T}  \otimes \bG^\dagger  \bG] 
\label{fun_gen}
\end{align}
The solution to the general  $\bPhi$ immediately follows from (\ref{fun_gen}).

\noindent
{\bf Solution for General} $\bPhi$: The optimum  $\bPhi$ solving (\ref{opt_Phi}) for a general non-diagonal $\bPhi$ is a reshaped (vector to matrix) and scaled version of the dominant eigenvector, $\bv_{\M}$, corresponding to the largest eigenvalue of $\bM$ in (\ref{fun_gen}): 
 \begin{equation}
 \bPhi_{opt} = \sqrt{\nis} \mathrm{reshape}(\bv_{\M} )    \ ; \ \bv_{\M}^\dagger \bv_{\M}  = 1  \ . \label{Phi_opt_gen} 
 \end{equation}
 
 The general $\bPhi$ design represents an advanced RIS in which
 \begin{wrapfigure}{l}{0.15\textwidth}
 \vspace{-2mm}
\includegraphics[width=0.15\textwidth]{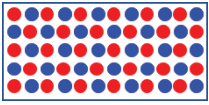}
\vspace{-8mm}
\caption{\footnotesize{\sl A potential realization of the general $\bPhi$. The blue elements collect the incoming signal $\by_i$ and the red elements emit the transformed signal $\by_o$. }}
\label{fig:ris_gen}
\vspace{-3mm}
\end{wrapfigure} 
the incoming signal vector $\by_i$ undergoes a general linear transformation (rather than an element-wise transformation) before it is emitted by the RIS: $\by_o = \bPhi \by_i$.   Fig.~\ref{fig:ris_gen} illustrates one potential realization of the general $\bPhi$. The RIS consists of an array of interleaved elements. The blue elements collect the incoming signal $\by_i$ and the red elements emit the transformed signal $\by_o$. 
\subsection{Low-Complexity Phase-Only Diagonal $\bPhi$ Design (LC-PH)}
\label{sec:lc}
For comparison, we also consider a low-complexity design for phase-only optimization of a diagonal $\bPhi$ proposed in \cite{ris_lc:20}.  First define the following vectors corresponding to $\bg_i$ and $\bh_i$ representing the absolute values of the elements:
\begin{align}
\bgtil_i & = \left [|g_{1,i}|, |g_{2,i}|, \cdots, |g_{\nr,i}| \right ]^{\T}  \nonumber \\
\bhtil^\dagger_i & = \left [ |h_{i,1}|, |h_{i,2}|, \cdots, |h_{i,\nt}| \right ]  \ , \ i = 1, 2, \cdots, \nis \ .
\label{ghtil}
\end{align}
Let $\bPhi_{lc} = \diag(e^{j \phi_{lc,1}}, e^{j \phi_{lc,2}}, \cdots, e^{j\phi_{lc,\nis}})$ denote the low-complexity phase-only diagonal $\bPhi$. The optimized choices for the phases at each of the RIS elements are given by \cite{ris_lc:20}
\begin{align}
\phi^{\HH}_{lc,i} & = \cos^{-1} \left( \frac{ \Re (\bhtil_i^\dagger \bh_i)}{\|\bh_i\| \| \bhtil_i\|} \right ) \nonumber \\
\phi^{\GG}_{lc,i} & = \cos^{-1} \left( \frac{ \Re (\bgtil_i^\dagger \bg_i)}{\|\bg_i\| \| \bgtil_i\|} \right ) \nonumber \\
\phi_{lc,i} & = -(\phi^{\HH}_{lc,i} + \phi^{\GG}_{lc,i})  \ , \ i = 1, 2, \cdots, \nis \label{phi_lc}
\end{align}
where $\Re(\cdot)$ denotes the real part and $\| \bx \| = \sqrt{\bx^\dagger \bx}$.  Note that $\bPhi_{lc}$ satisfies the constraint in (\ref{opt_Phi}).
\subsection{Baseline Non-Engineered Diagonal $\bPhi$ (RAND)}
\label{sec:baseline}
For a baseline comparison, we also consider choices of $\bPhi$ which are not engineered or optimized, as may occur in reflections from a physical object of the same size as the RIS. We consider two choices for baseline diagonal $\bPhi_{bl}$: i) diagonal entries correspond to random phases \{$e^{j\phi_{bs,i}}$ \}, ii) diagonal entries correspond to random phases and amplitude \{$ \cCN(0,1)$\}. The first choice automatically satisfies the constraint in (\ref{opt_Phi}) and in the second case the entries are normalized to satisfy (\ref{ris_const}). 

\begin{figure}[hbt]
\begin{tabular}{cc}
\hspace{-3mm}
\includegraphics[width=0.24\textwidth]{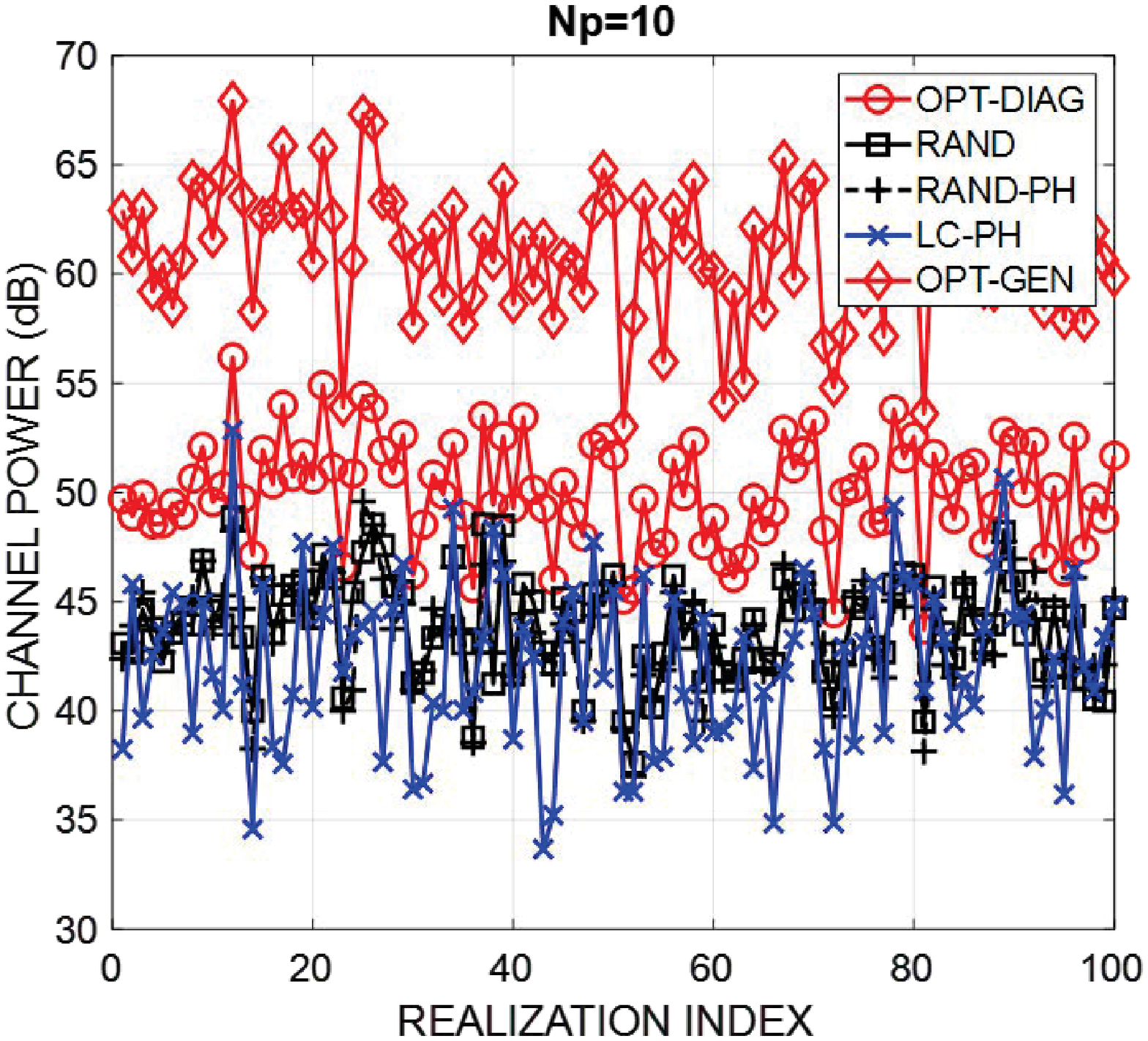} &  
\hspace{-5mm} \includegraphics[width=0.24\textwidth]{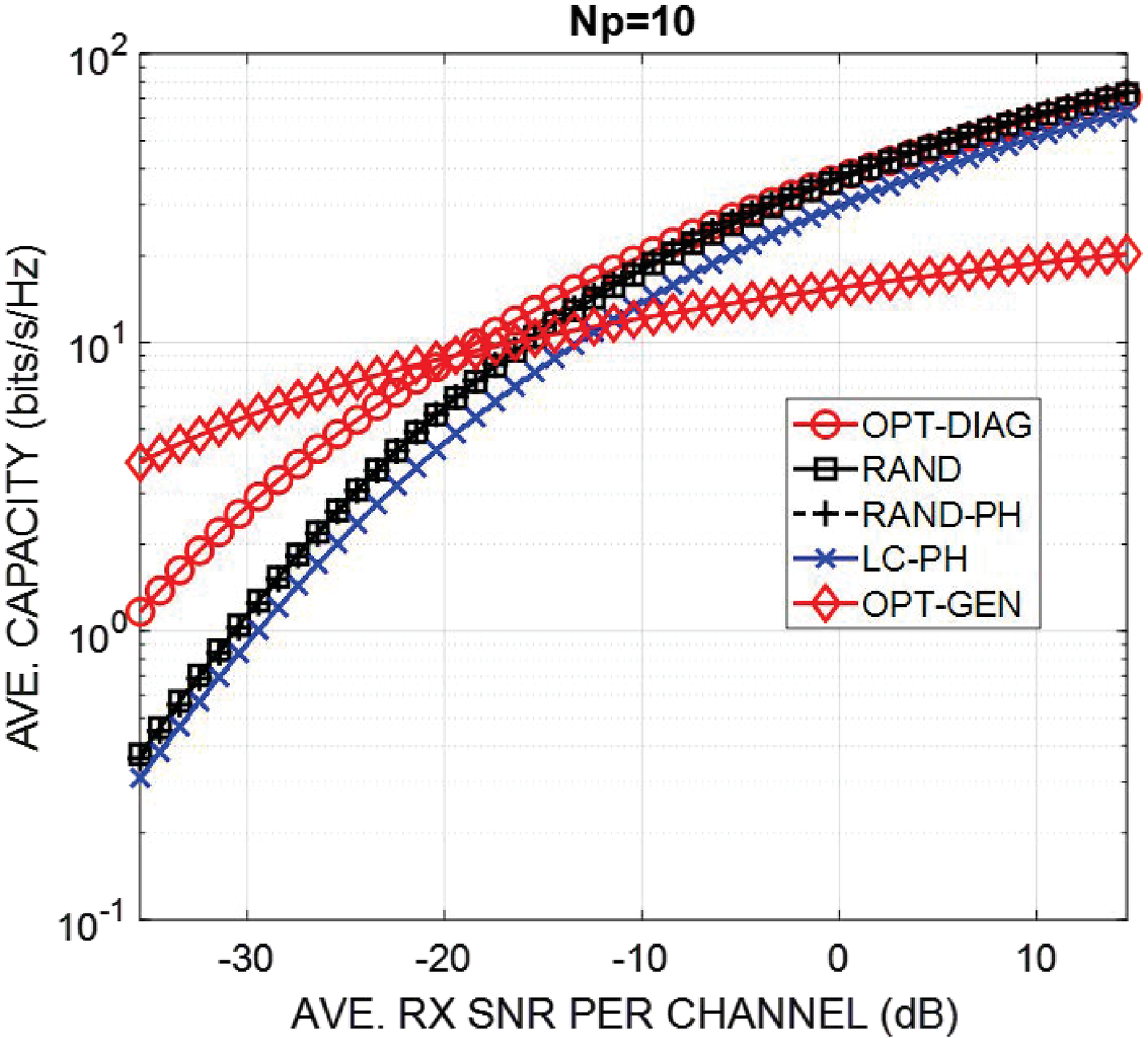} \\ 
\hspace{-3mm} \footnotesize{(a)}  & \hspace{-5mm}  \footnotesize{(b)} \\
\hspace{-3mm} \includegraphics[width=0.24\textwidth]{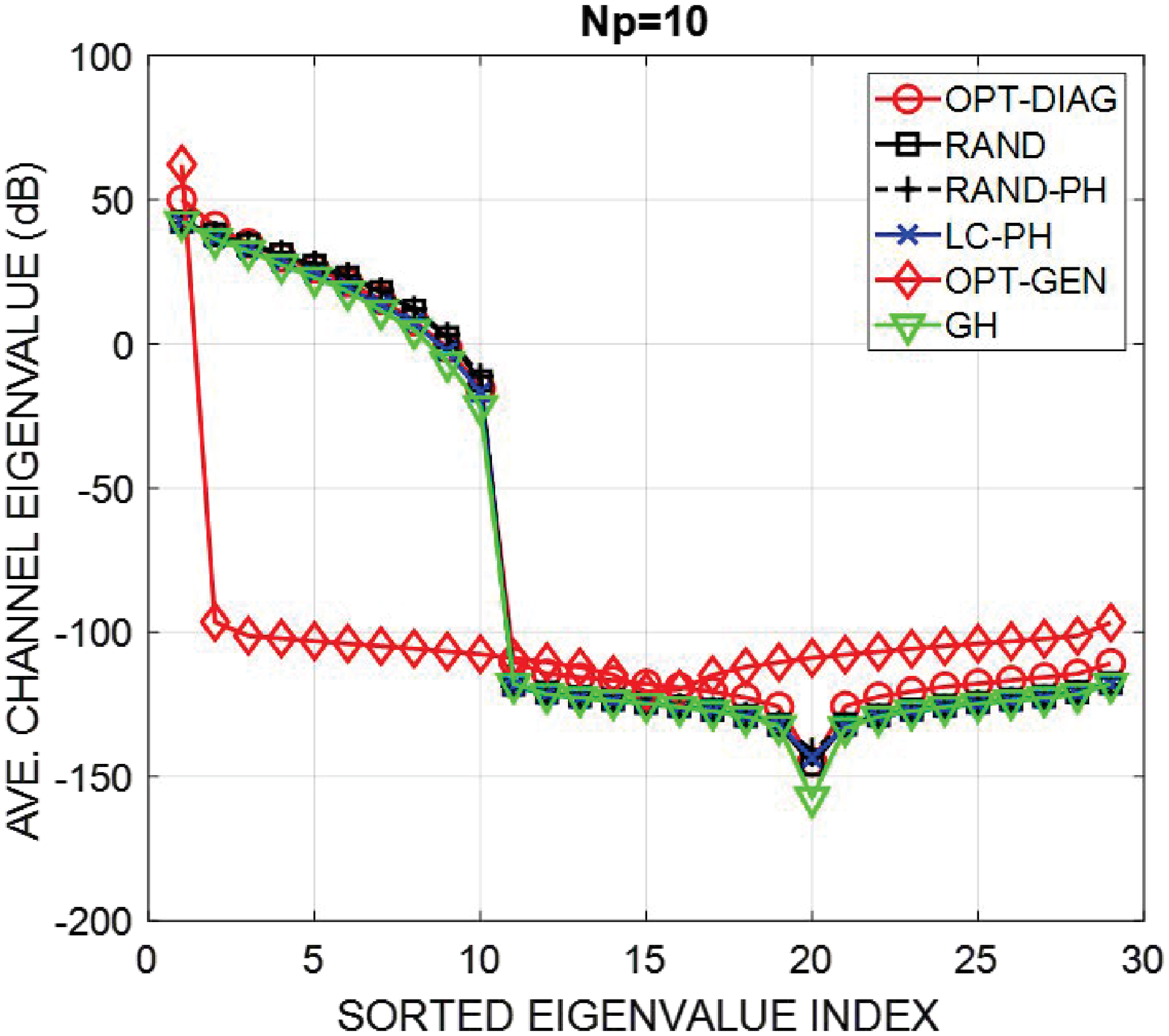} & 
 \hspace{-5mm} \includegraphics[width=0.24\textwidth]{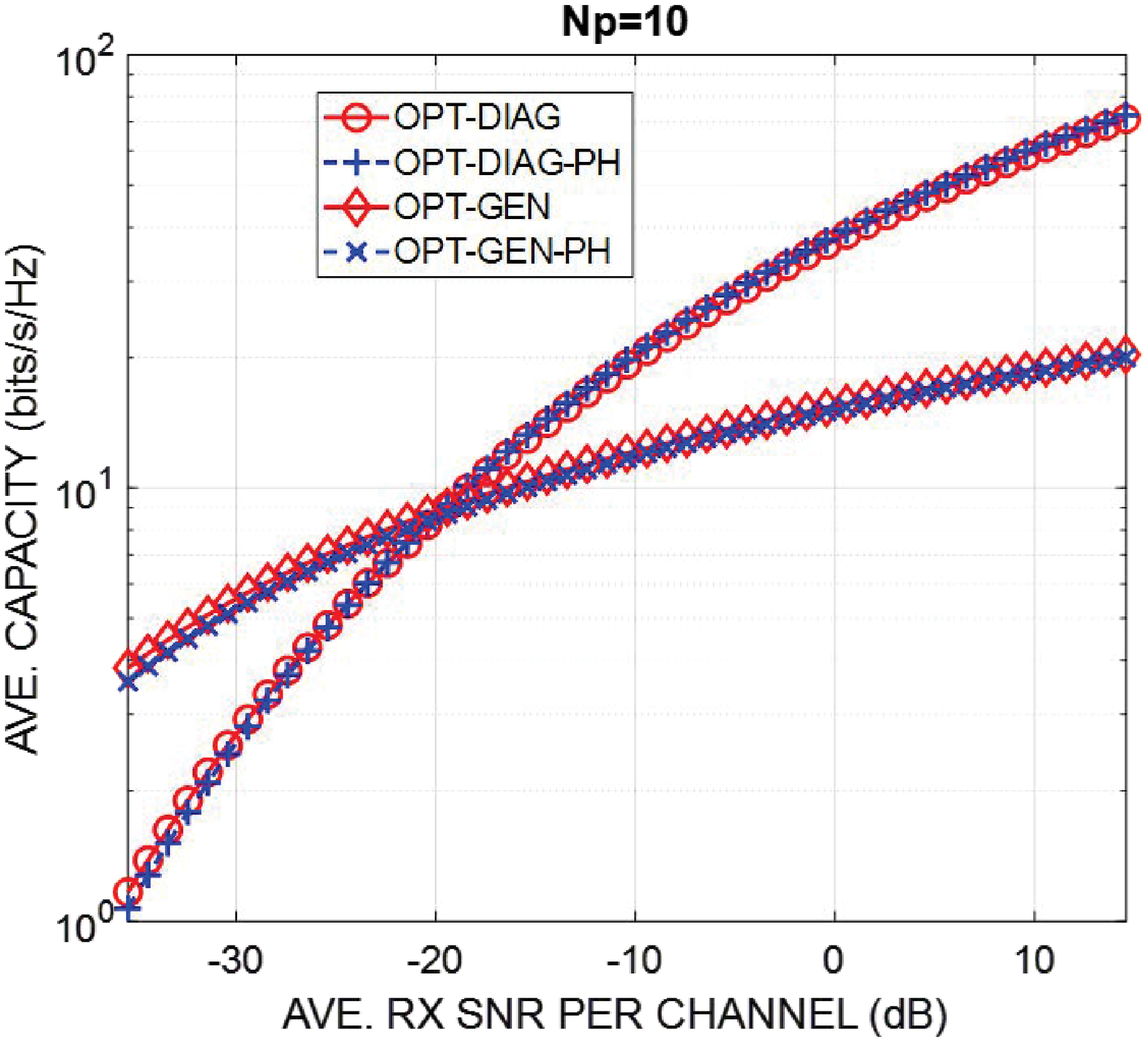} \\ 
\hspace{-3mm} \footnotesize{(c)} & \hspace{-5mm}   \footnotesize{(d)}
\end{tabular}
\caption{\footnotesize{\sl Performance of different $\bPhi$ designs for {\bf NLoS sparse multipath}  ($N_p = 10$): (a) channel power; (b) link capacity; (c) channel eigenvalues; (d) capacity comparison between OPT and OPT-PH designs.}}
\label{fig:nlos_sp}
\vspace{-3mm}
\end{figure}

\section{Numerical Results and Discussion}
\label{sec:res}
In this section, we illustrate the performance of the various RIS designs through numerical results.  We consider a mmW TX-RIS-RX link operating at 28 GHz with the TX, RIS and RX consisting of 6-inch linear apertures with $\nt = \nr = \nis = 29$ critically spaced elements. We consider both NLoS and NLoS + LoS propagation environments for $\bH$ and $\bG$. In each case, we consider a sparse multipath scenario with $N_p = N_{p}^{\GG} = N_{p}^{\HH} = 10$ paths and a richer multipath scenario with $N_p = N_p^{\GG} = N_p^{\HH} = 100$ paths. The path powers are normalized so that the average channel powers, $E[\tr(\bH^\dagger \bH)]$ and $E[\tr(\bG^\dagger \bG)]$, are constant. We compare the performance of five $\bPhi$ designs: i) $\bPhi_{opt}$ with diagonal elements (OPT),   ii) general $\bPhi_{opt}$ (OPT-GEN), iii) low-complexity phase-only design $\bPhi_{lc}$ (LC-PH), iv) baseline design $\bPhi_{bl}$ with random complex entries (RAND), and v) baseline design  $\bPhi_{bl}$ with random phases (RAND-PH). For the optimum designs, we also consider the phase-only versions, where we replace $\bPhi_{opt}$ entries with their respective phase values (OPT-PH and OPT-GEN-PH).  We assume perfect CSI, knowledge of $\bH$, $\bG$ and  $\bF = \bG \bPhi \bH$, at the TX, RX and RIS and compare the capacity of the five $\bPhi$ designs, computed through optimal power allocation (waterfilling algorithm) over the channel eigenvalues. The capacity is averaged over 100 channel realizations where each realization corresponds to an independent realization of complex path gains, AoAs and AoDs for both $\bH$ and $\bG$. The AoAs and AoDs are randomly and uniformly distributed over the entire angular spreads. We note that the maximum number of spatial channels in the TX-RIS-RX link is given by $N_{ch}=\min(\nr,\nt,\nis) = 29$.   Thus, the  average RX SNR per channel is $\snr_{ch} = P_{ave} E[\sigma^2_{\bF}]/N_{ch}$ (see (\ref{rx_pow})).

\subsection{NLoS Propagation }
\label{sec:nlos}
\begin{figure}[thb]
\begin{tabular}{cc}
\hspace{-3mm}
\includegraphics[width=0.24\textwidth]{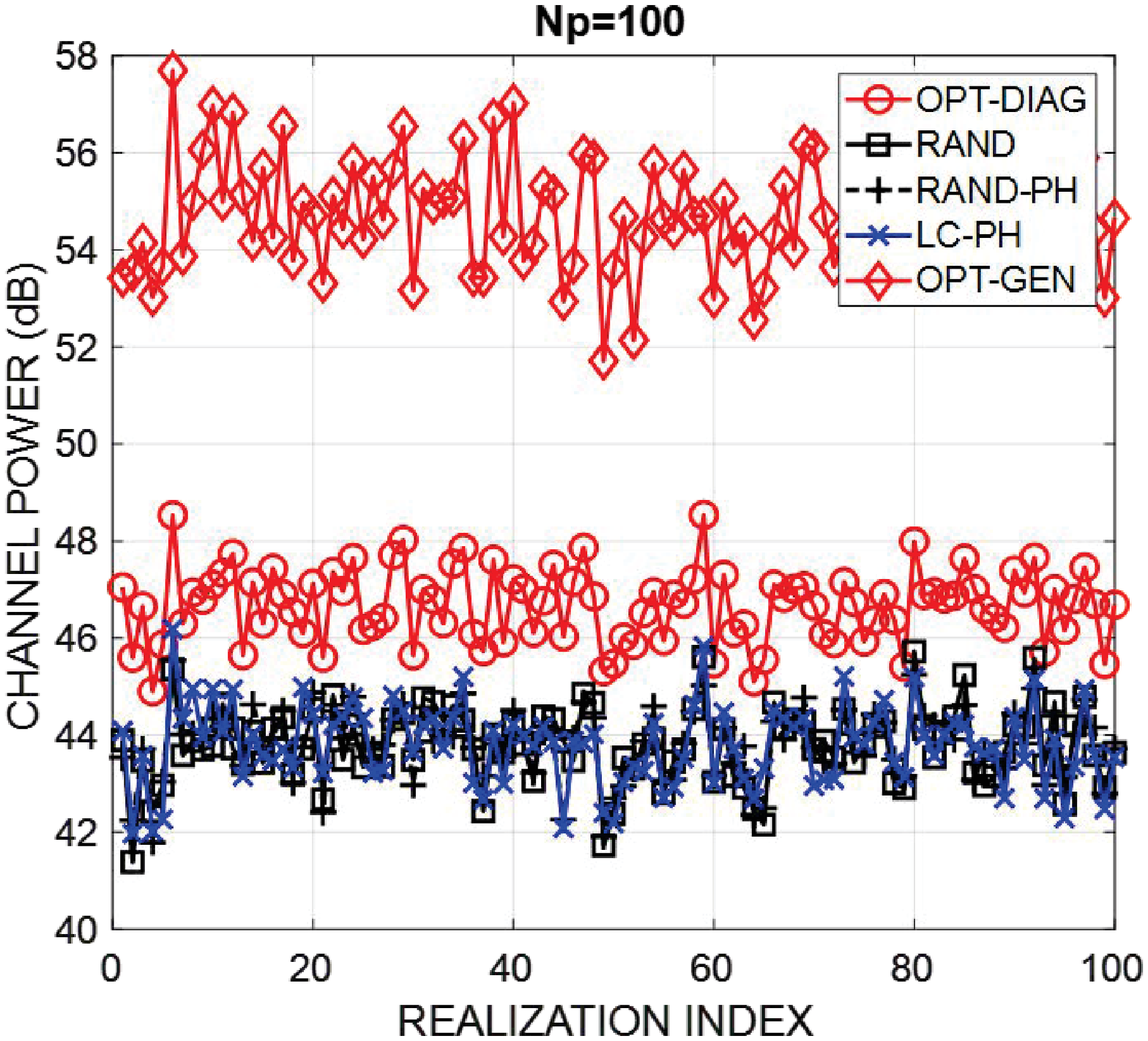} &  
\hspace{-5mm} \includegraphics[width=0.24\textwidth]{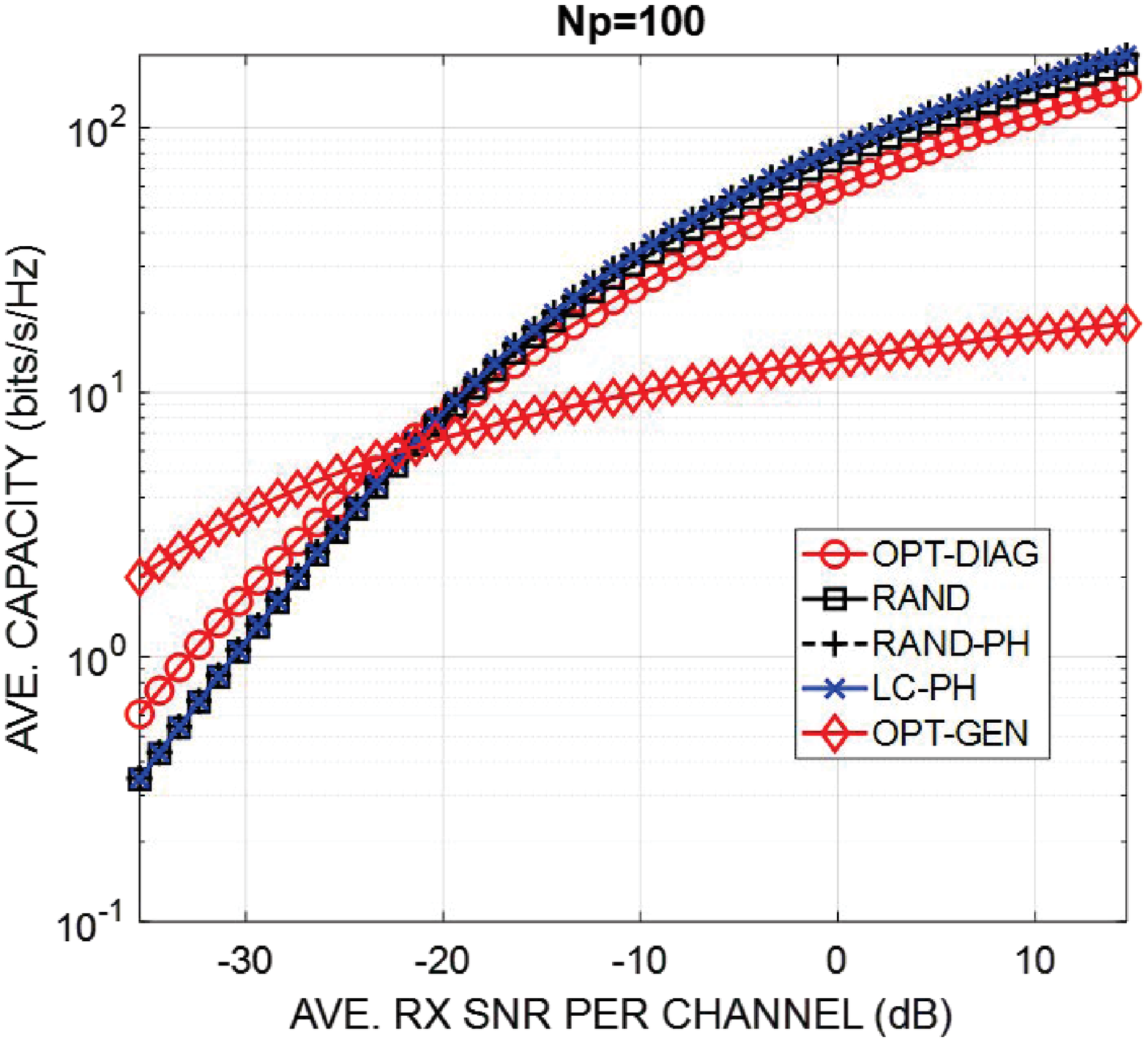} \\ 
\hspace{-3mm} \footnotesize{(a)}  & \hspace{-5mm}  \footnotesize{(b)} \\
\hspace{-3mm} \includegraphics[width=0.24\textwidth]{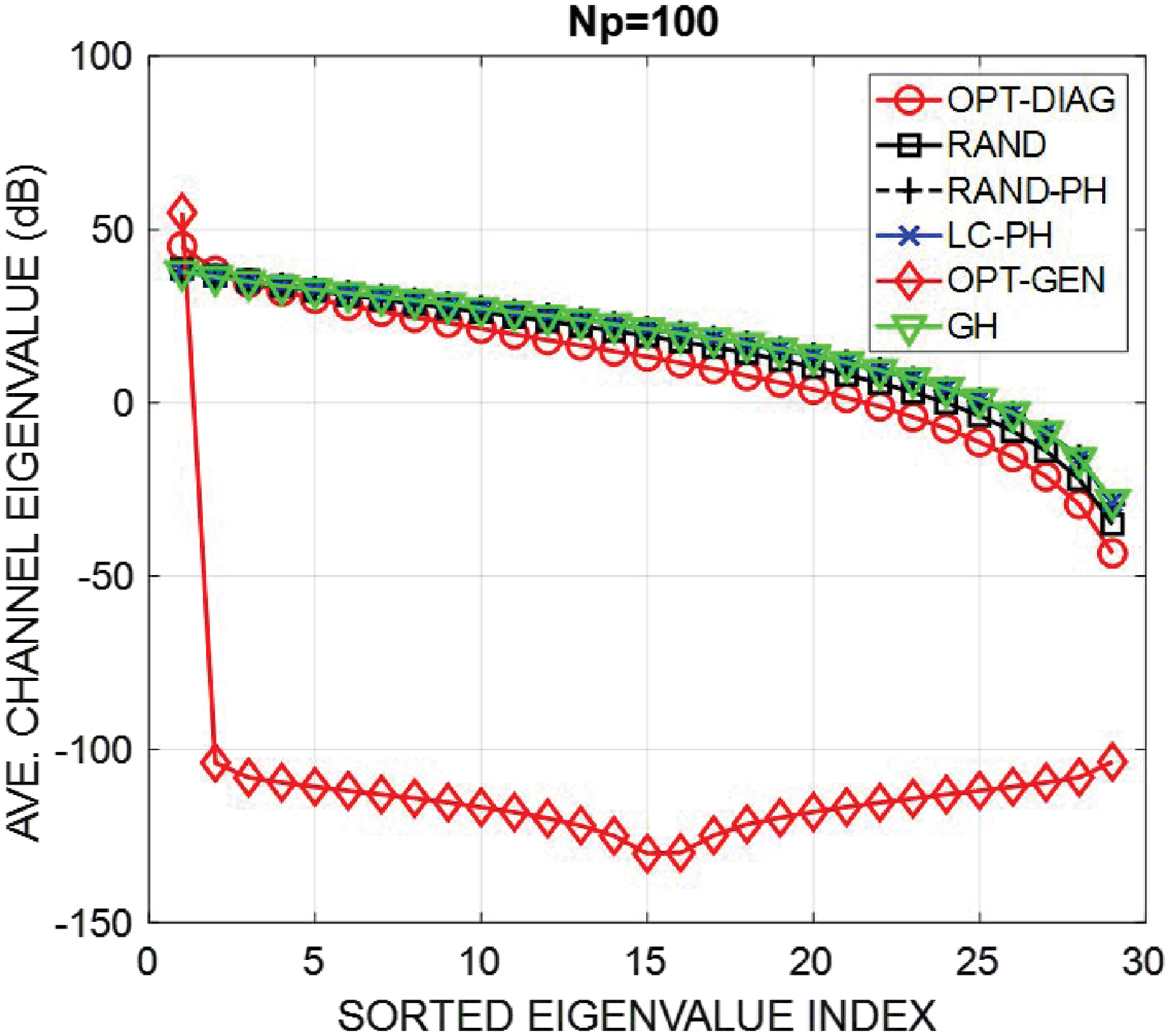} & 
 \hspace{-5mm} \includegraphics[width=0.24\textwidth]{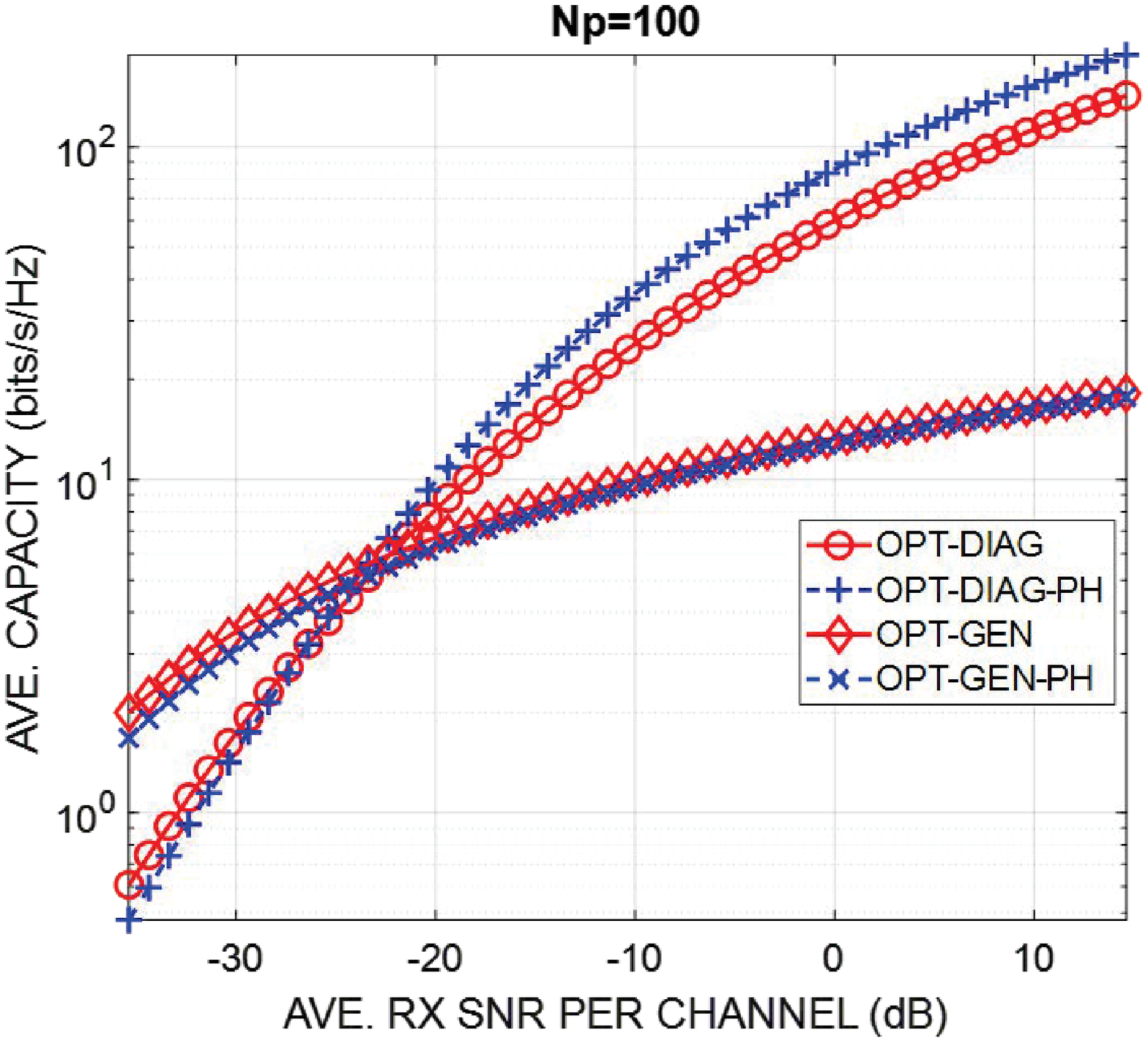} \\ 
\hspace{-3mm} \footnotesize{(c)} & \hspace{-5mm}   \footnotesize{(d)}
\end{tabular}
\caption{\footnotesize{\sl Performance of different $\bPhi$ designs for {\bf NLoS rich multipath}  ($N_p = 100$): (a) channel power; (b) link capacity; (c) channel eigenvalues; (d) capacity comparison between OPT and OPT-PH designs.}}
\label{fig:nlos_rich}
\vspace{-3mm}
\end{figure}
In this section, we consider NLoS propagation for both $\bH$ and $\bG$ where all paths have equal powers. Fig.~\ref{fig:nlos_sp} shows the results for the sparse multipath environment $N_p=10$. Fig.~\ref{fig:nlos_sp}(a) plots the channel powers $\sigma^2_{\bF} = \tr(\bF^\dagger \bF)$ for different $\bPhi$ designs; it is clear that both RAND and LC-PH have comparable channel power, OPT has higher power and OPT-GEN has the highest power. Fig.~\ref{fig:nlos_sp}(b) plots the average channel capacity (bits/s/Hz) as a function of the average RX SNR per channel, $\snr_{ch}$, for different $\bPhi$ choices. As evident, the capacities of RAND and RAND-PH are nearly identical and slightly higher than LC-PH. The OPT-DIAG design delivers higher capacity than RAND/LC at lower  values of $\snr_{ch}$. On the other hand, OPT-GEN delivers significant capacity gains at lower $\snr_{ch}$ values while incurring a significant loss at higher $\snr_{ch}$. The capacity behavior is governed by the eigenvalues of $\bF^\dagger \bF$ and Fig.~\ref{fig:nlos_sp}(c) plots the average values of sorted (ordered) channel eigenvalues. It is evident that RAND, LC-PH and OPT-DIAG have comparable eigenvalues, with dominant one being larger for OPT-DIAG reflecting its better capacity at lower $\snr_{ch}$. The ``GH" plot corresponds to $\Phi = \bI_{\nis}$.  Note that the 10 largest eigenvalues are significantly larger reflecting $N_p=10$.  The dominant channel eigenvalue for OPT-GEN is largest at the cost of lower values for the remaining eigenvalues reflecting its significantly higher capacity at lower $\snr_{ch}$ and corresponding loss at higher $\snr_{ch}$. Fig.~\ref{fig:nlos_sp}(d) compares the capacities of OPT designs with the corresponding phase-only versions and the two capacities are nearly identical. Thus, the phase-only versions of $\bPhi_{opt}$, which are simpler to implement, may not incur any loss in  performance.

Fig.~\ref{fig:nlos_rich} shows the corresponding results for NLoS propagation for richer multipath $N_p=100$.  Channel powers in Fig.~\ref{fig:nlos_rich}(a) show less variation due to richer multipath diversity. Interestingly, while the average power for RAND and LC-PH is about the same as in sparse multipath (Fig.~\ref{fig:nlos_sp}(a)), the average power for OPT designs is lower. The impact on capacity in Fig.~\ref{fig:nlos_rich}(b) is that all designs have a higher capacity at high $\snr_{ch}$, compared to the sparse scenario, whereas the low-SNR gain in capacity of OPT designs over RAND and LC-PH is reduced due to their lower average powers. The relative trends in channel eigenvalues in Fig.~\ref{fig:nlos_rich}(c) are similar to that in the spare NLoS scenario, except that in this case  the eigenvalues for all of the $N_{ch} = 29$ spatial eigen-channels for RAND, LC-PH and  OPT-DIAG are significant due to richer multipath. The concentration of channel power into essentially a single eigenvalue for OPT-GEN is the same as in Fig.~\ref{fig:nlos_sp}(c).  Interestingly, the high-SNR capacity of phase-only version of OPT-DIAG in Fig.~\ref{fig:nlos_rich}(d) is slightly higher in this case compared to sparse multipath.
\begin{figure}[hbt]
\vspace{-3mm}
\begin{tabular}{cc}
\hspace{-3mm}
\includegraphics[width=0.24\textwidth]{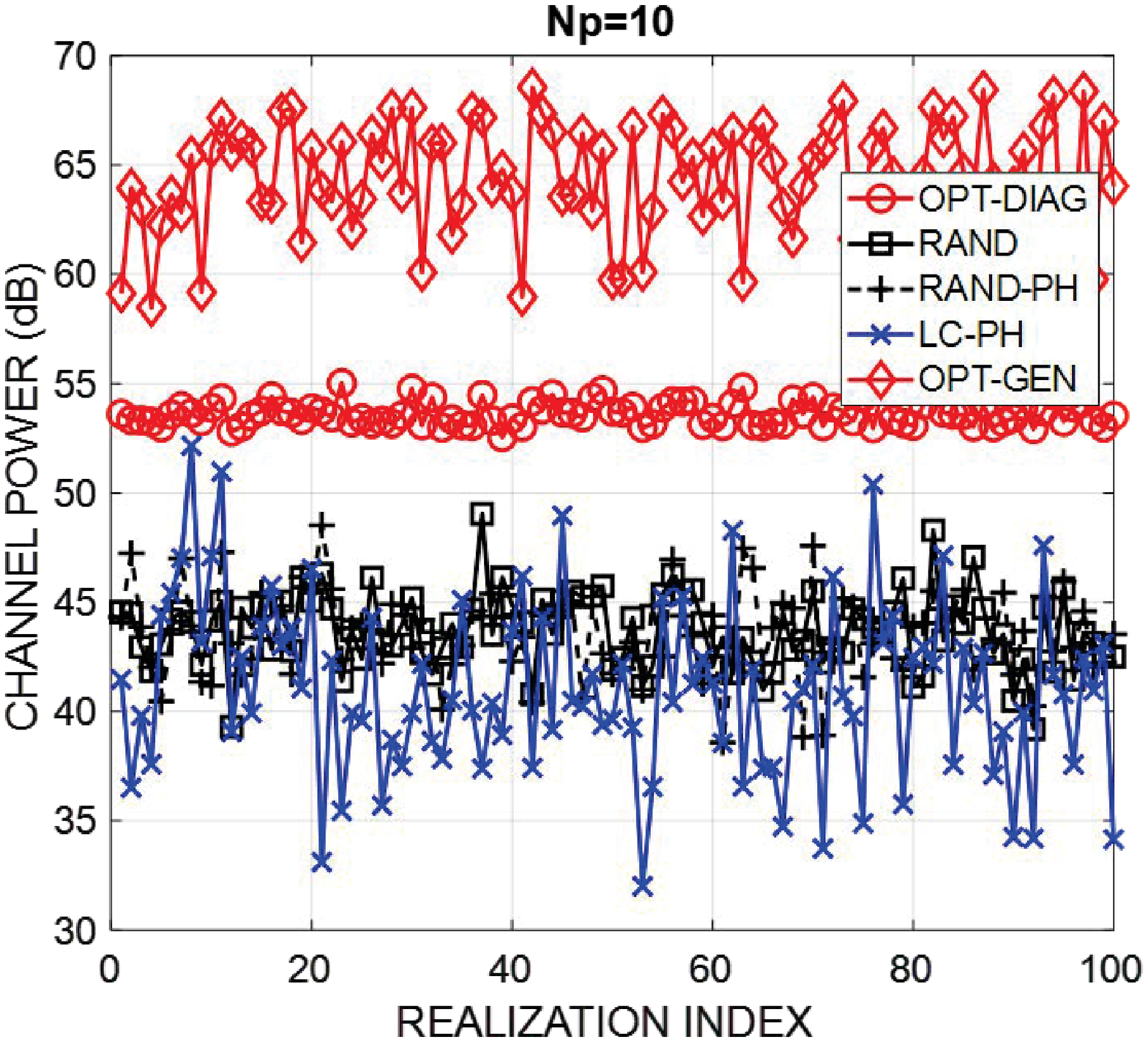} &  
\hspace{-5mm} \includegraphics[width=0.24\textwidth]{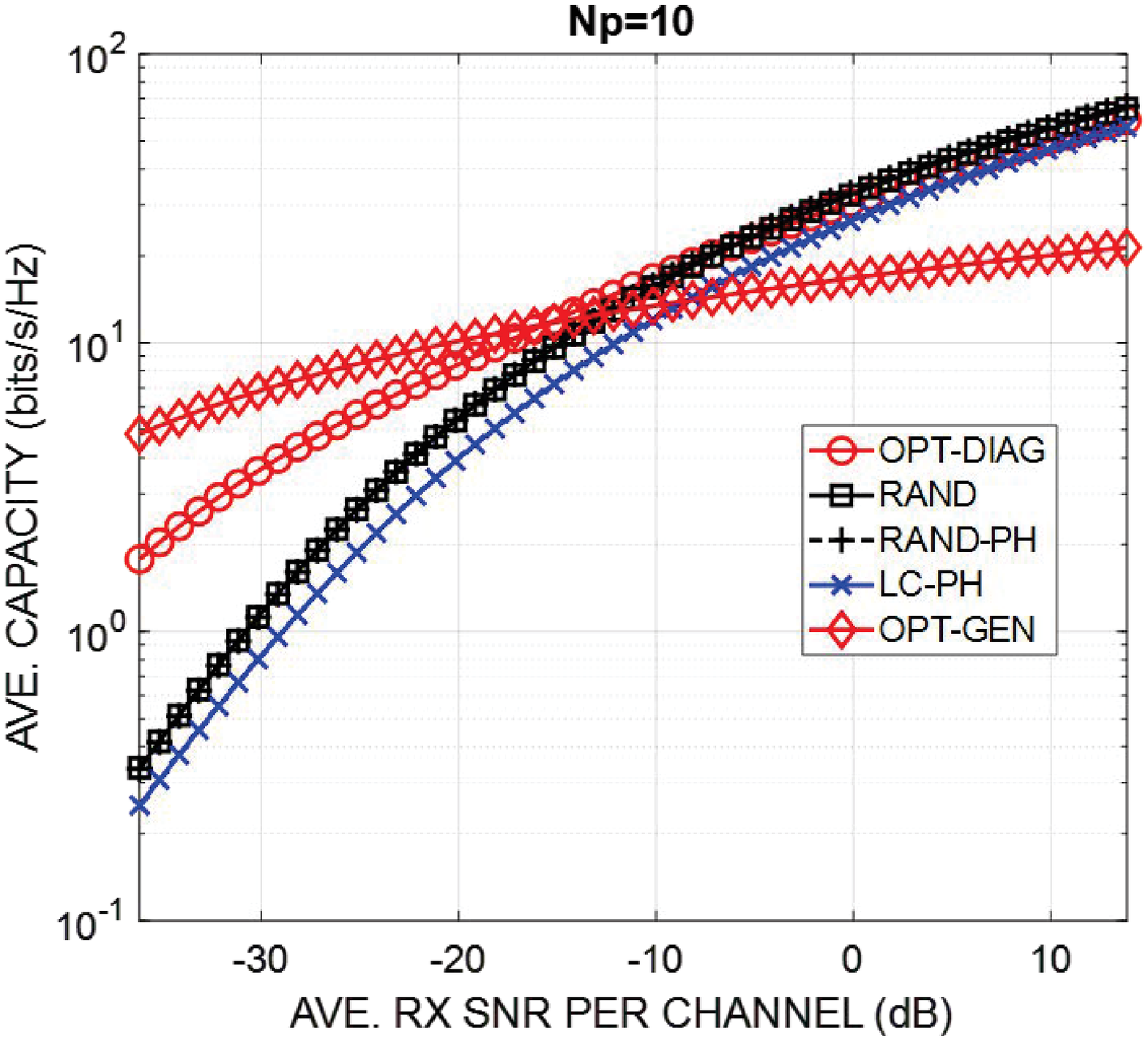} \\ 
\hspace{-3mm} \footnotesize{(a)}  & \hspace{-5mm}  \footnotesize{(b)} \\
\hspace{-3mm} \includegraphics[width=0.24\textwidth]{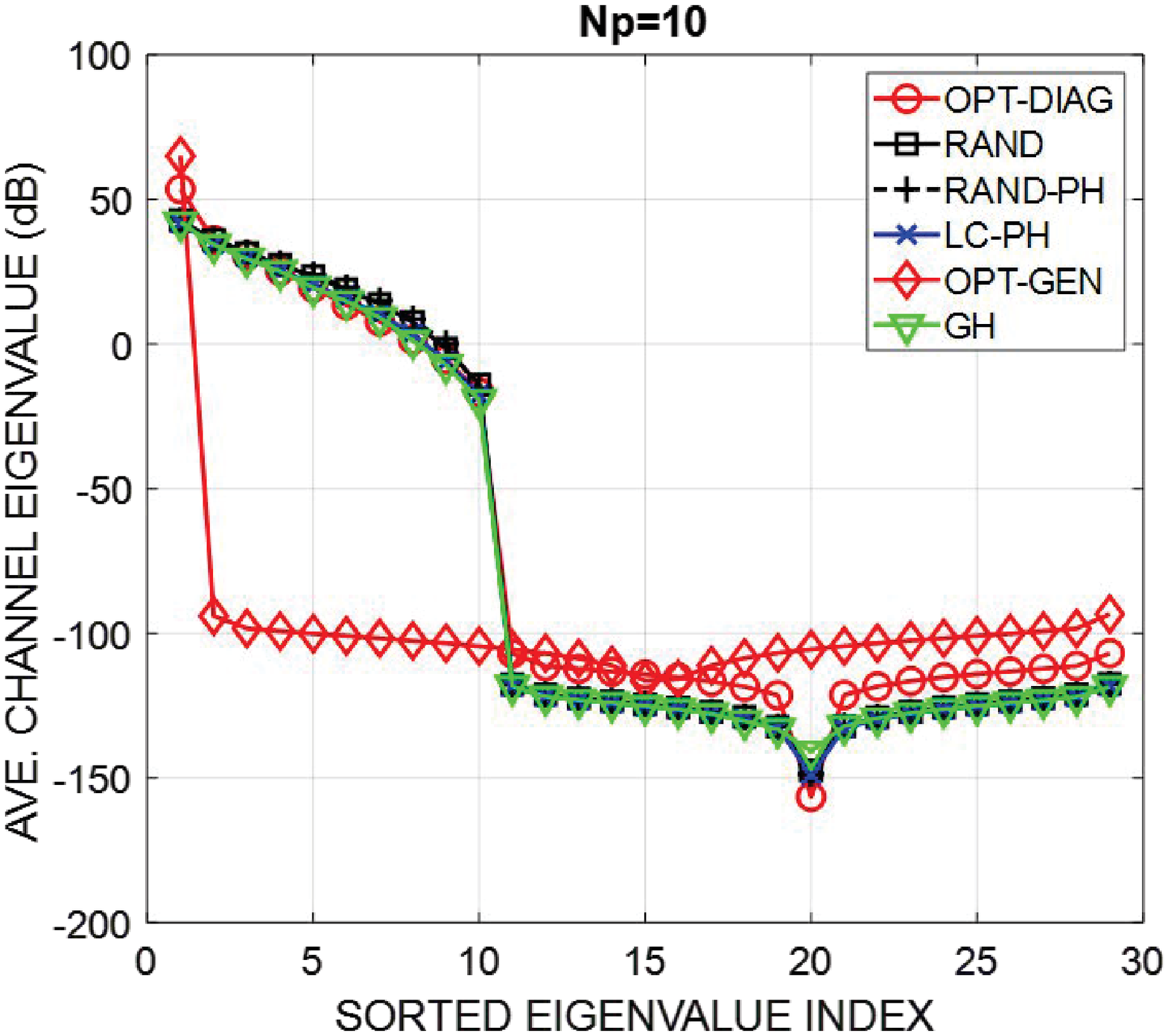} & 
 \hspace{-5mm} \includegraphics[width=0.24\textwidth]{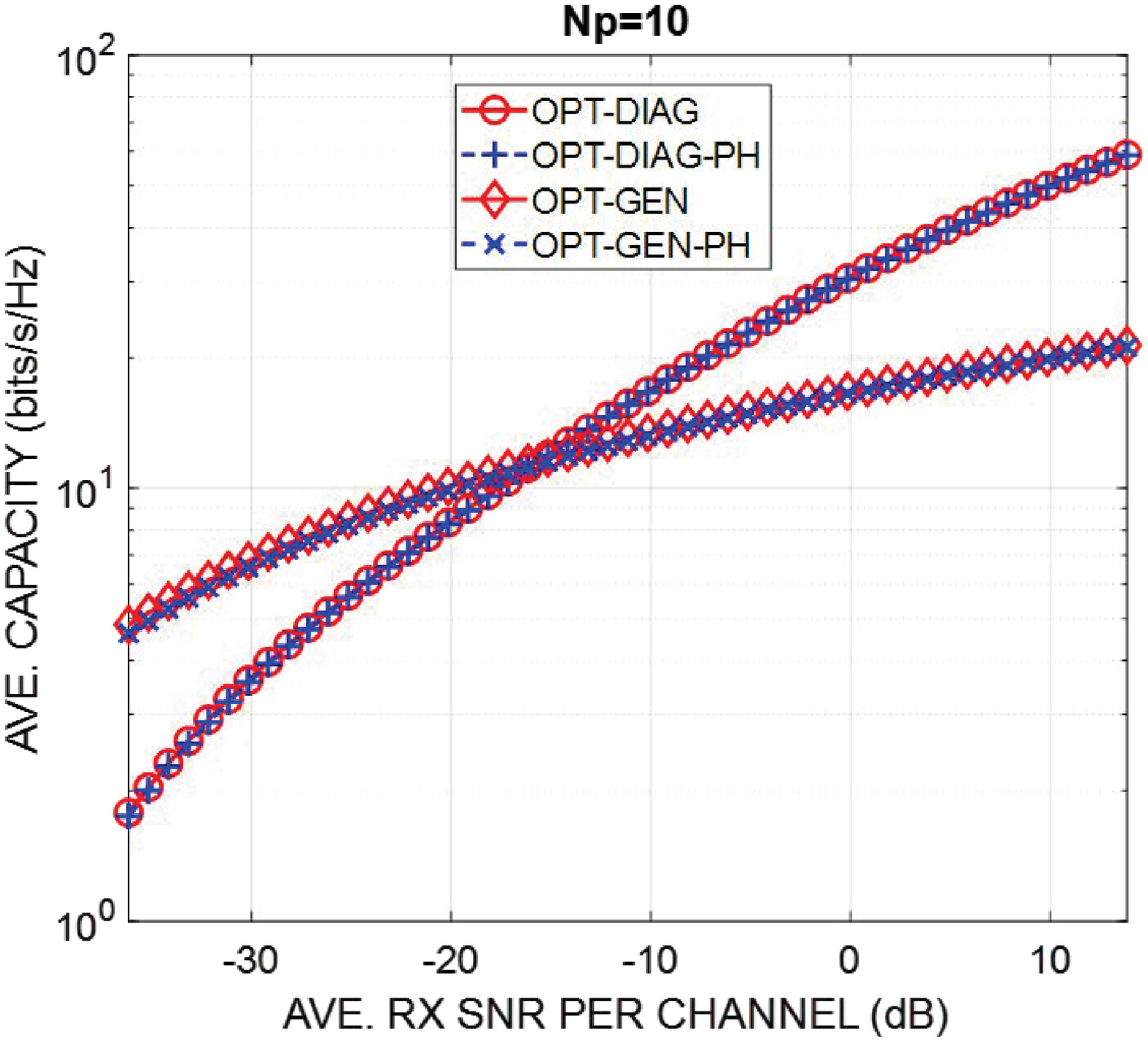} \\ 
\hspace{-3mm} \footnotesize{(c)} & \hspace{-5mm}   \footnotesize{(d)}
\end{tabular}
\caption{\footnotesize{\sl Performance of different $\bPhi$ designs for {\bf LoS + NLoS sparse multipath} ($N_p = 10$): (a) channel power; (b) link capacity; (c) channel eigenvalues; (d)  capacity comparison between OPT and OPT-PH designs.}}
\label{fig:los_sp}
\vspace{-3mm}
\end{figure}

\subsection{NLoS + LoS Propagation}
\label{sec:los}
\begin{figure}[hbt]
\begin{tabular}{cc}
\hspace{-3mm}
\includegraphics[width=0.24\textwidth]{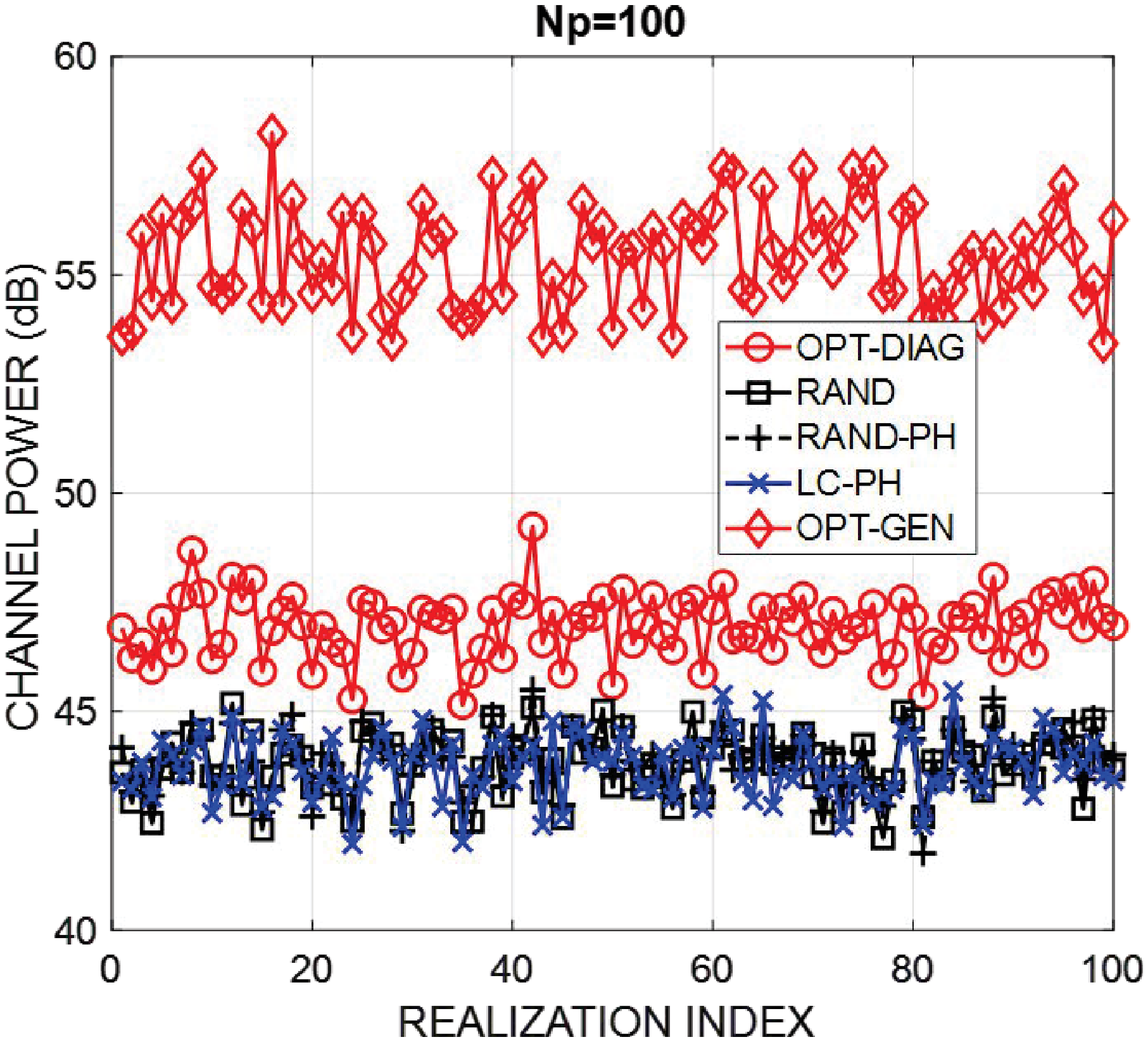} &  
\hspace{-5mm} \includegraphics[width=0.24\textwidth]{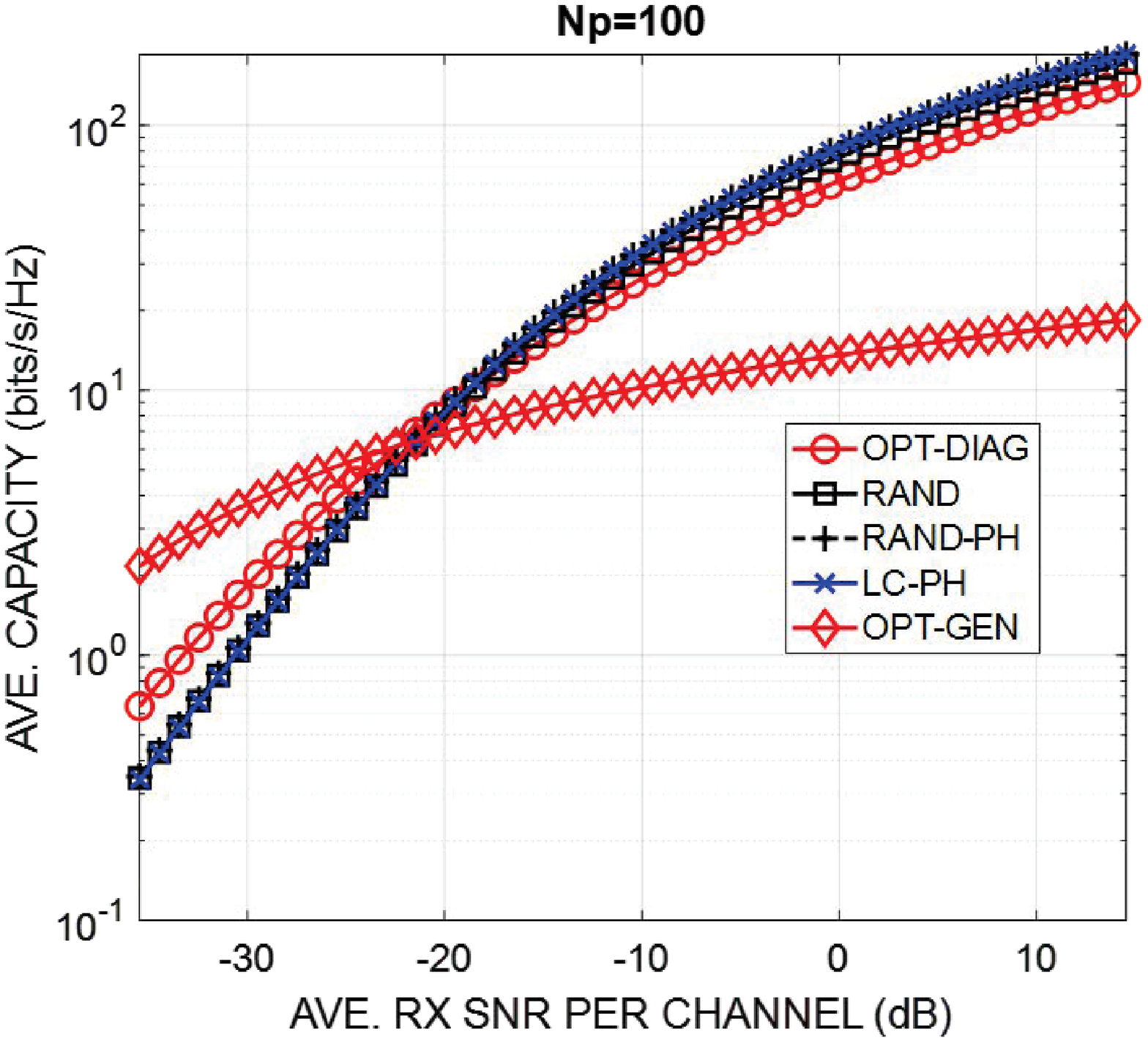} \\ 
\hspace{-3mm} \footnotesize{(a)}  & \hspace{-5mm}  \footnotesize{(b)} \\
\hspace{-3mm} \includegraphics[width=0.24\textwidth]{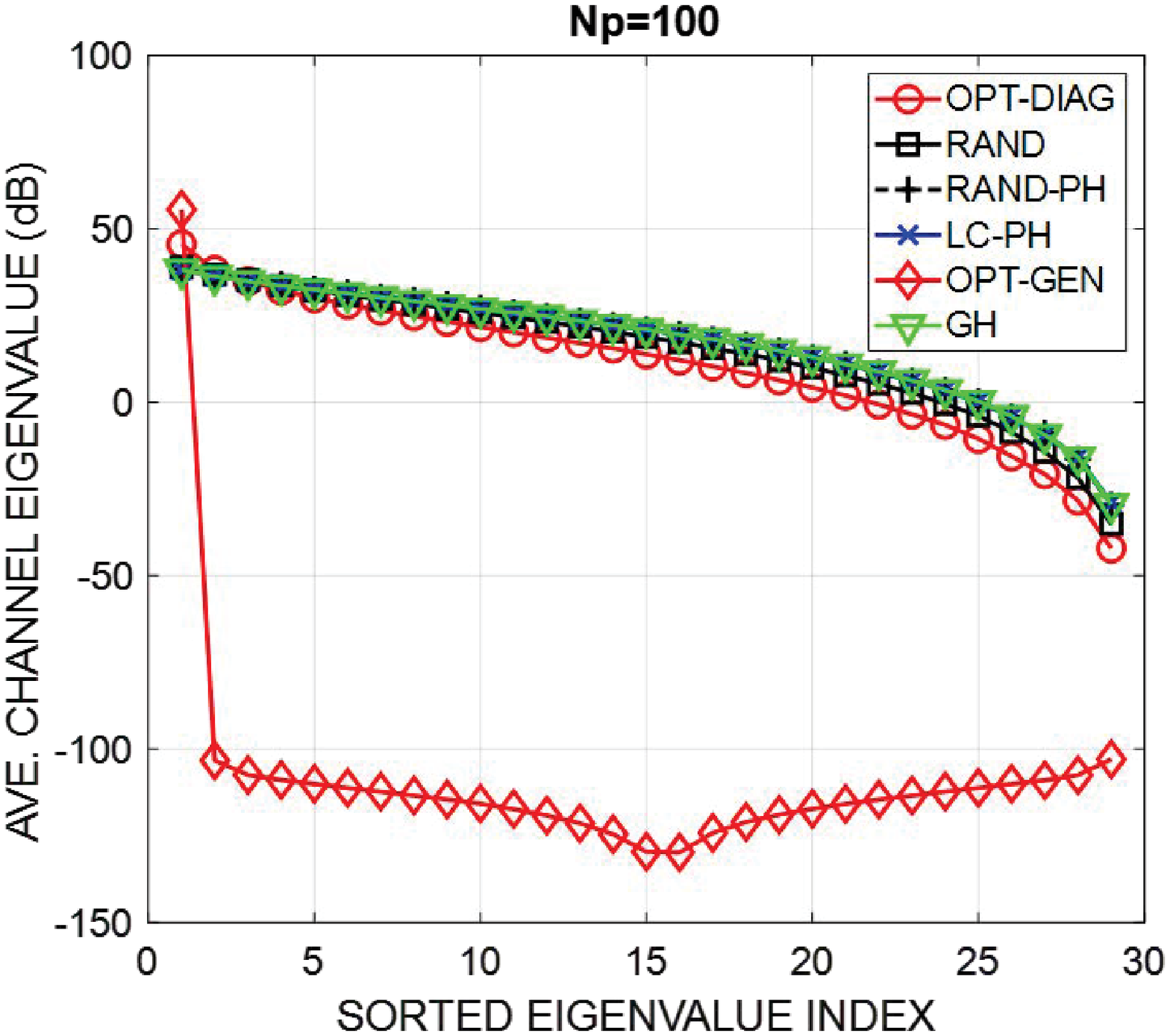} & 
 \hspace{-5mm} \includegraphics[width=0.24\textwidth]{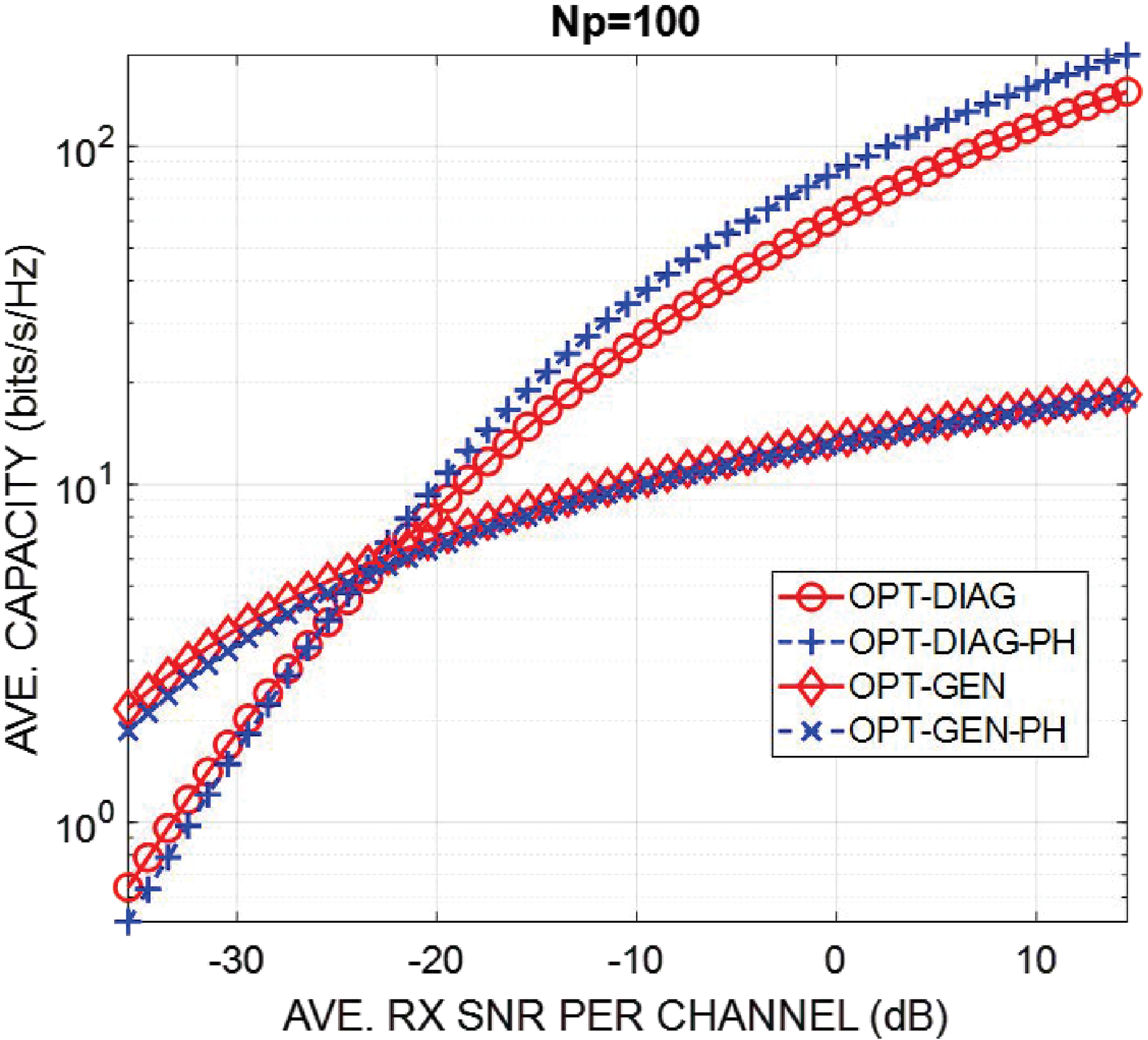} \\ 
\hspace{-1mm} \footnotesize{(c)} & \hspace{-5mm}   \footnotesize{(d)}
\end{tabular}
\caption{\footnotesize{\sl Performance of different $\bPhi$ designs for {\bf LoS + NLoS rich multipath} ($N_p = 10$): (a) channel power; (b) link capacity; (c) channel eigenvalues; (d) capacity comparison between OPT and OPT-PH designs.}}
\label{fig:los_rich}
\end{figure}
We now present results for propagation with both  LoS and NLoS paths in both $\bH$ and $\bG$,  with the  rest of the parameters the same. The LoS path  introduces only a phase shift and has 10dB stronger power than the remaining equal-power NLoS paths with complex Gaussian amplitudes. Fig.~\ref{fig:los_sp} plots the results for the sparse scenario ($N_p=10$) and Fig.~\ref{fig:los_rich} for the richer multipath case ($N_p=100$).  The general trends in this case are similar to the NLoS plots in Figs.~\ref{fig:nlos_sp}-\ref{fig:nlos_rich}. One notable difference is the reduction in variation of channel power for OPT-DIAG due the presence of the LoS component. In terms of capacity, the performance of RAND and LC-PH designs is very comparable to the NLoS only case. On the other hand, while the OPT designs show similar high-SNR capacity, the low-SNR capacity is  higher due to the LoS component.

\subsection{Impact of the Size of RIS}
\label{sec:ris}
\begin{figure}[bht]
\begin{tabular}{cc}
\hspace{-1mm}
\includegraphics[width=0.24\textwidth]{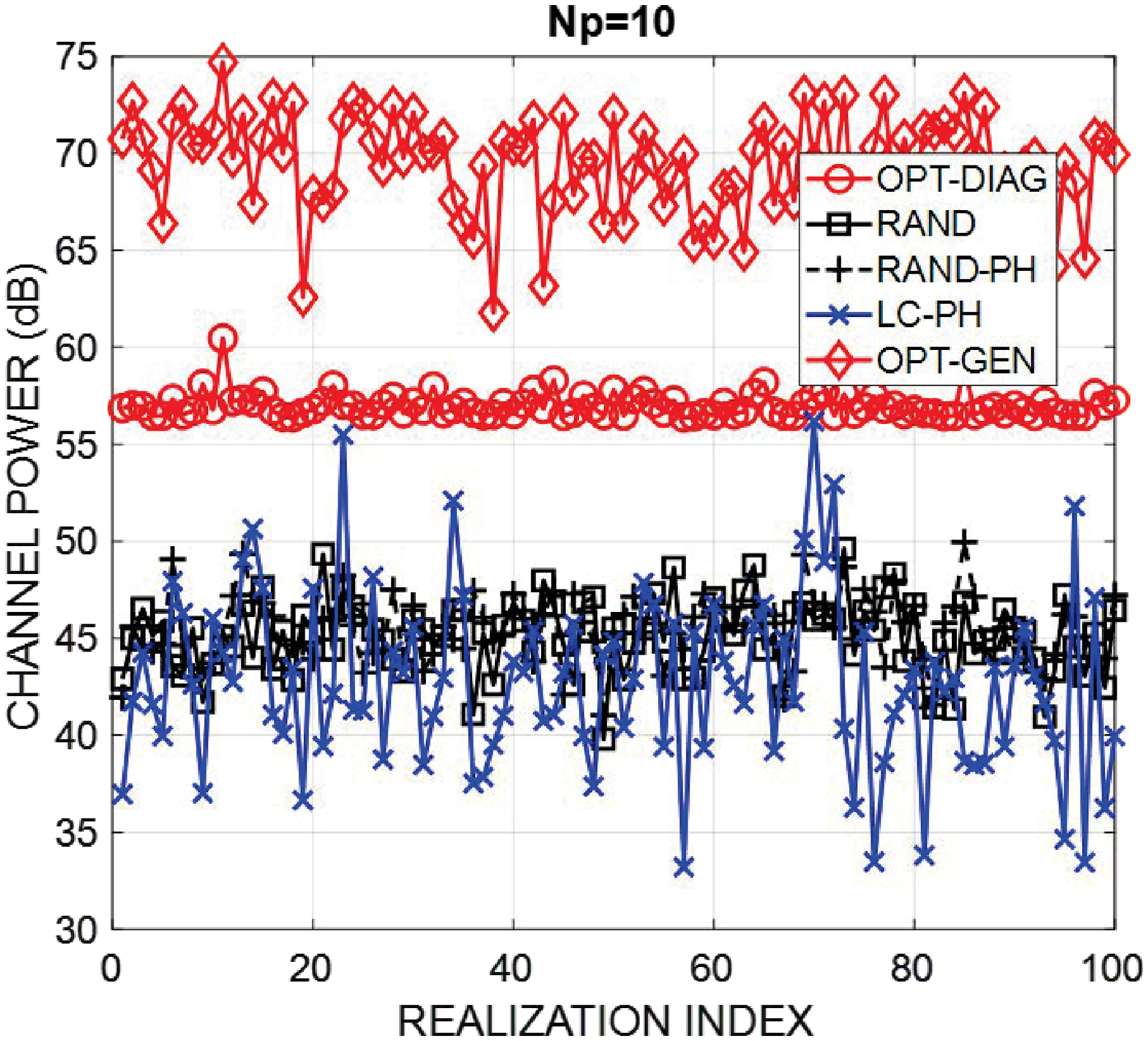} &  
\hspace{-5mm} \includegraphics[width=0.24\textwidth]{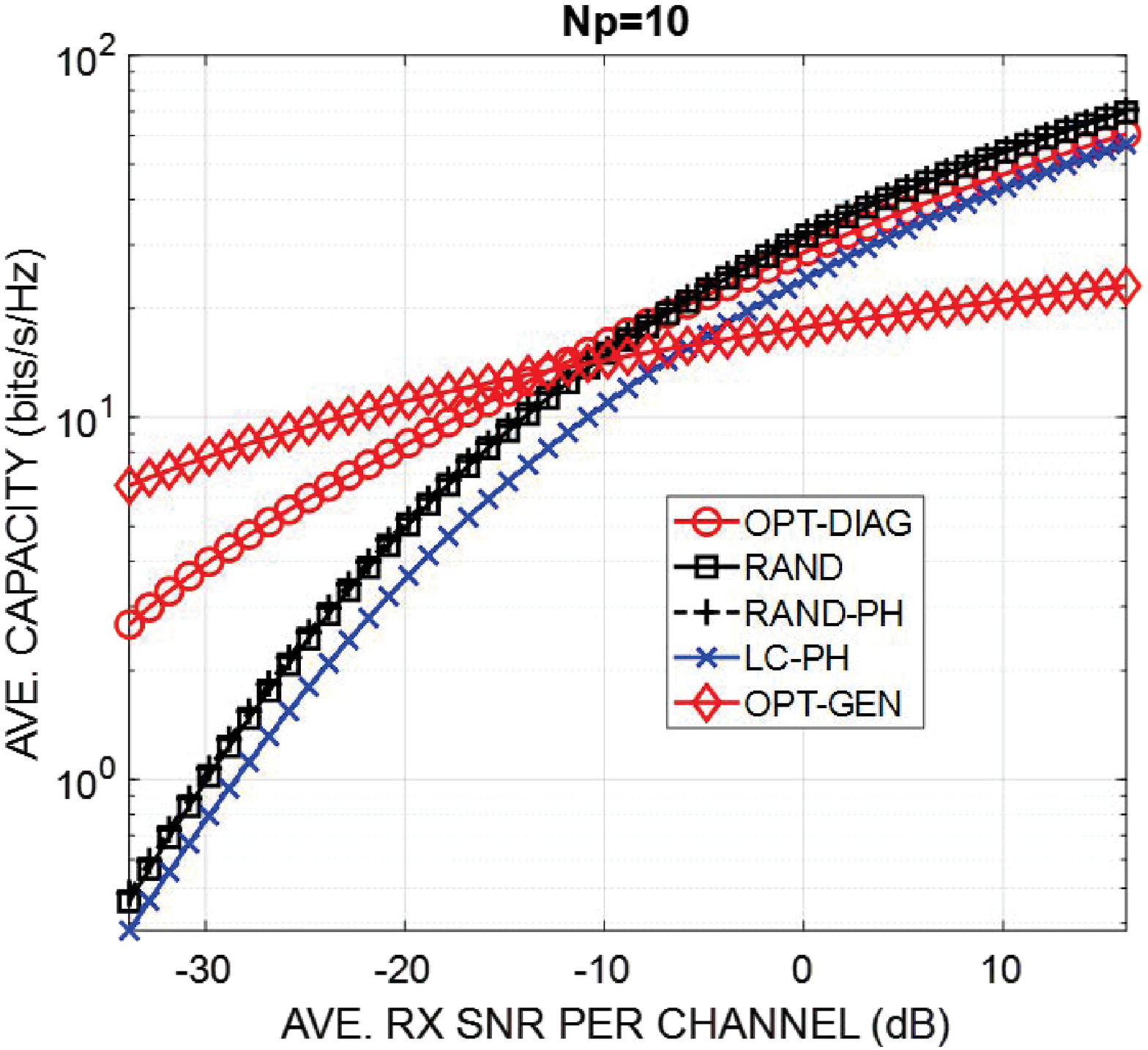} \\ 
\hspace{-3mm} \footnotesize{(a)}  & \hspace{-5mm}  \footnotesize{(b)} \\
\hspace{-3mm} \includegraphics[width=0.24\textwidth]{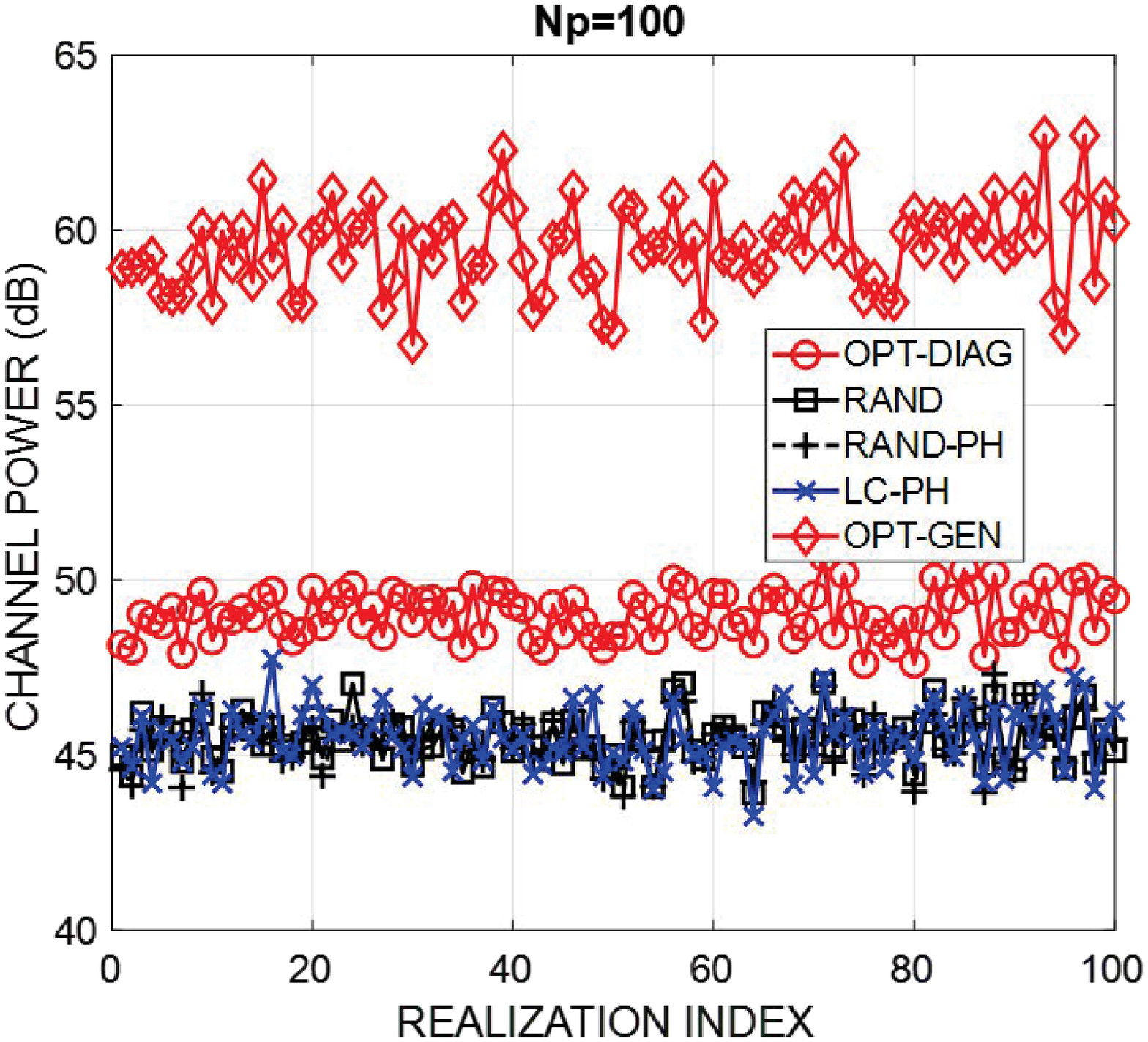} & 
 \hspace{-5mm} \includegraphics[width=0.24\textwidth]{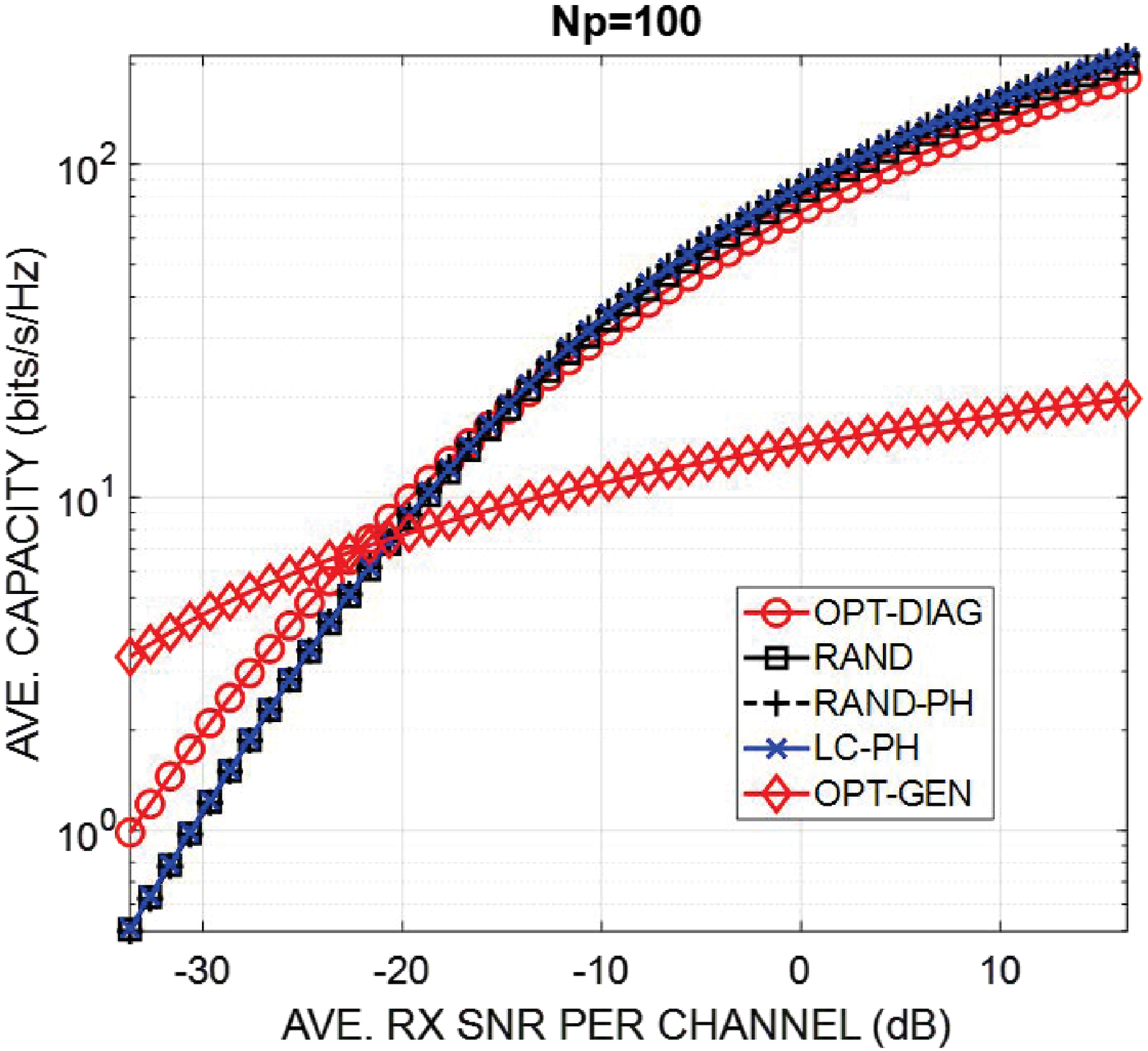} \\ 
\hspace{-3mm} \footnotesize{(c)} & \hspace{-5mm}   \footnotesize{(d)}
\end{tabular}
\caption{\footnotesize{\sl Performance of different $\bPhi$ designs for {\bf LoS + NLoS multipath} and {\bf larger RIS} ($\nis = 43$): (a)-(b) sparse multpath - (a) channel power, (b) link capacity; (c)-(d) rich multipath -  (c) channel power, (d) link capacity.}}
\label{fig:los_ris}
\vspace{-3mm}
\end{figure}
We now present results to illustrate the impact of a larger RIS in the LoS + NLoS environment in Sec.~\ref{sec:los}.  The RIS has $\nis = 43$ elements corresponding to a 9" aperture. Figs.~\ref{fig:los_ris}(a) and (b) plot the channel powers and capacity for the sparse scenario and Figs.~\ref{fig:los_ris}(c) and (d) show the corresponding results for richer multipath. The trends are the same as for the smaller RIS case with the overall effect of uniformly increasing channel powers and capacity due to the larger RIS.

\section{Conclusion}
\label{sec:conc}
In this paper, we have proposed a framework for designing optimized RISs based on maximizing the average power collected by the receiver. The initial results on capacity indicate that the resulting designs can deliver significant capacity gains in the low-SNR regime compared to an un-engineered RIS (RAND) or the low-complexity (LC-PH) design. The optimized diagonal design delivers competitive performance at high-SNR and the general optimized design delivers even higher gains at low SNRs at the cost of high-SNR performance. One observation, that is consistent across all scenarios, is that the RAND design performs as well as the LC-PH design and actually gives slightly higher capacity in sparse multipath; this may be due to the approximate nature of the LC-PH design. Furthermore, the low-SNR capacity gains of the OPT designs are generally  higher in sparse multipath and/or in the presence of a LoS component.  Our results also indicate that the phase-only versions of the optimized designs, which are simpler to implement, incur no loss in performance and can even yield slightly higher capacity at high SNRs. This is because the optimization criterion does not directly reflect capacity. The framework proposed in this paper opens up new directions for research including: probability of error analysis of proposed designs; extensions to and time- and frequency-selective channels \cite{sayeed:handbook08}; extensions to multiuser network scenarios; impact of estimated CSI; and physical implementation of the general $\bPhi$ design.

\bibliographystyle{IEEEtran}


\end{document}